\documentclass[prl,showpacs,twocolumn,
               amsmath,amssymb,superscriptaddress,
               floatfix,nofootinbib]{revtex4-1}

\usepackage[utf8]{inputenc}
\usepackage{amsmath,amssymb,amsbsy,amsfonts}
\usepackage{bm}
\usepackage{times}

\usepackage{graphicx}
\usepackage{float}
\usepackage{morefloats}
\usepackage{rotating}

\usepackage[caption=false]{subfig}

\usepackage{booktabs}
\usepackage{multirow}
\usepackage{tabularx}
\usepackage{adjustbox}
\usepackage{nicefrac}
\usepackage{slashed}
\usepackage{verbatim}
\usepackage[dvipsnames]{xcolor}
\usepackage{ragged2e}

\usepackage{hyperref}
\hypersetup{
	colorlinks	=true,
	urlcolor	=blue,
	linkcolor	=blue,
	citecolor	=blue,
	pdftitle	={},
	pdfauthor	={Jia-Ming Xie, Jun-Xu Lu, Li-Sheng Geng},
	pdfsubject	={Two-pole structures demystified: chiral dynamics at work}
	}


\begin{document}

\title{Chiral Evolution and Femtoscopic Signatures of the $K_1(1270)$ Resonance}

\author{Jia-Ming Xie}
\affiliation{School of Physics, Beihang University, Beijing 102206, China}
\affiliation{Department of Physics, Graduate School of Science, The University of Tokyo, Tokyo 113-0033, Japan}

\author{Zhi-Wei Liu}
\affiliation{School of Physics, Beihang University, Beijing 102206, China}

\author{Jun-Xu Lu}
\email[Corresponding author: ]{ljxwohool@buaa.edu.cn}
\affiliation{School of Physics, Beihang University, Beijing 102206, China}

\author{Haozhao Liang}
\email[Corresponding author: ]{haozhao.liang@phys.s.u-tokyo.ac.jp}
\affiliation{Department of Physics, Graduate School of Science, The University of Tokyo, Tokyo 113-0033, Japan}
\affiliation{Quark Nuclear Science Institute, The University of Tokyo, Tokyo 113-0033, Japan}
\affiliation{RIKEN Center for Interdisciplinary Theoretical and Mathematical Sciences, Wako 351-0198, Japan}

\author{Raquel Molina}
\email[Corresponding author: ]{Raquel.Molina@ific.uv.es}
\affiliation{
Departamento de F\'isica Te\'orica and IFIC, 
Centro Mixto Universidad de Valencia-CSIC, 
Parc Cient\'ific UV, C/ Catedr\'atico Jos\'e Beltr\'an, 2, 
46980 Paterna, Spain
}

\author{Li-Sheng Geng}
\email[Corresponding author: ]{lisheng.geng@buaa.edu.cn}
\affiliation{Sino-French Carbon Neutrality Research Center, \'{E}cole Centrale de P\'{e}kin/School
of General Engineering, Beihang University, Beijing 100191, China}
\affiliation{School of
Physics,  Beihang University, Beijing 102206, China}
\affiliation{Peng Huanwu Collaborative Center for Research and Education, Beihang University, Beijing 100191, China}
\affiliation{Beijing Key Laboratory of Advanced Nuclear Materials and Physics, Beihang University, Beijing 102206, China }
\affiliation{Southern Center for Nuclear-Science Theory (SCNT), Institute of Modern Physics, Chinese Academy of Sciences, Huizhou 516000, China}

\begin{abstract}
We present a comprehensive study of the axial-vector resonance $K_1(1270)$ within the unitarized chiral perturbation theory, focusing on its two-pole structure and manifestation in femtoscopic observables.  
By considering the dominant $\rho K$ and $K^*\pi$ coupled channels, we reproduce the well-established double-pole structure and trace the chiral evolution of both poles as functions of the pion mass, using the vector-meson mass trajectories fitted to lattice-QCD data and experimental values.  
The lower pole, dominantly coupled to $K^*\pi$, evolves from an above-threshold resonance to a virtual or bound state with increasing pion mass. In comparison, the higher pole, dominantly coupled to $\rho K$, moves downward in energy, reflecting the strengthening of the chiral attraction.  
The influence of the finite vector-meson widths is systematically examined, showing that their inclusion smooths the pole trajectories without altering their qualitative behavior.
Furthermore, femtoscopic CFs are calculated for all relevant vector-pseudoscalar channels in both charged sectors.  
The results exhibit distinct resonance and bound-state features consistent with the two-pole dynamics.  
The weak impact of higher channels, such as $\omega \bar{K}$, $\bar{K}^*\eta$, and $\phi\bar{K}$, confirms that the simplified two-channel treatment captures the essential dynamics of the $K_1(1270)$ resonance.  
This study demonstrates that combining chiral extrapolation and femtoscopic correlation analyses provides a powerful and complementary framework for connecting lattice-QCD calculations, chiral effective theory, and experimental measurements, offering new insights into the molecular nature and chiral origin of the $K_1(1270)$ resonance.
\end{abstract}
%\pacs{13.75.Gx, 13.75.Jz,12.39.Fe}

\maketitle

%%%%%%%%%%%%%%%%%%%%%%%%%%%%%%%
\section{Introduction}
\label{sec:intro}
Understanding the internal structure of hadronic resonances remains one of the most challenging problems in nonperturbative QCD~\cite{Brambilla:2010cs,Chen:2016qju,Richard:2016eis,Esposito:2016noz,Lebed:2016hpi,Hosaka:2016ypm,Guo:2017jvc,Olsen:2017bmm,Karliner:2017qhf,Liu:2019zoy,Mai:2022eur,Chen:2022asf,Liu:2024uxn}. 
Among them, the low-lying axial-vector mesons, such as $a_1(1260)$, $f_1(1285)$, and $K_1(1270)$~\cite{Lutz:2003fm,Roca:2005nm}, provide an essential probe into chiral symmetry breaking and the dynamics of vector-pseudoscalar~(VP) interactions. 
In the conventional quark model~\cite{Godfrey:1985xj}, these resonances are described as the $L=1$ orbital excitations of $q\bar{q}$ states, forming the $1^3P_1$ and $1^1P_1$ nonets. 
The strangeness partners $K_1(1270)$ and $K_1(1400)$ are usually interpreted as mixtures of the SU(3) eigenstates $K_{1A}$ $(C=+1)$ and $K_{1B}$ $(C=-1)$ due to SU(3)-flavor symmetry breaking.

However, certain long-standing experimental inconsistencies---such as reaction-dependent line shapes, channel-dependent branching ratios ($K^*\pi$ vs.\ $\rho K$), and interference with $K_1(1400)$---challenge this simple mixing picture~\cite{Geng:2006yb}. 
The coexistence of multiple peaks in diffractive and non-diffractive $Kp$ reactions suggests that the internal structure of $K_1(1270)$ is more complex than a single Breit-Wigner resonance~\cite{Mai:2022eur}.

The development of unitarized chiral perturbation theory~(UChPT) provides a compelling alternative~\cite{Weinberg:1990rz,Kaiser:1995eg,Kaiser:1996js,Oset:1997it,Oller:2000ma,Hyodo:2011ur,Oller:2019opk}. 
By unitarizing chiral VP interactions~\cite{Birse:1996hd,Roca:2005nm}, UChPT dynamically generates axial-vector states as hadronic molecules rather than conventional $q\bar{q}$ excitations. 
Notably, UChPT~\cite{Roca:2005nm} predicted two nearby poles associated with $K_1(1270)$, originating from the coupled-channel dynamics among $K^*\pi$, $\rho K$, $\omega K$, $K^*\eta$, and $\phi K$. 
The analysis of the high-statistics WA3 data on $K^-p \!\to\! K^-\pi^+\pi^-p$~\cite{Geng:2006yb} further supports the existence of two poles: a lower one around $1195 - 123i~\mathrm{MeV}$ dominantly coupling to $K^*\pi$, and a higher one around $1284 - 73i~\mathrm{MeV}$ mainly coupling to $\rho K$. 
This provided strong evidence that $K_1(1270)$ is a superposition of two dynamically generated states rather than a simple $K_{1A}$--$K_{1B}$ mixture.

Subsequent studies reinforced this picture. 
Wang et al.~\cite{Wang:2019mph} analyzed $D^0 \!\to\! \pi^+VP$ decays, showing that different final states manifest different poles---$K^*\pi$ the lower one, and $\rho K$ the higher. 
Later, their study of semileptonic $D^+ \!\to\! \nu e^+VP$ decays~\cite{Wang:2020pyy} confirmed this channel dependence. 
Dias et al.~\cite{Dias:2021upl} extended the analysis to $\bar{B} \!\to\! J/\psi\rho\bar{K}$ and $\bar{B} \!\to\! J/\psi\bar{K}^*\pi$, finding both poles in the line shapes of the $\rho K$ and $K^*\pi$ spectra. 
Roca et al.~\cite{Roca:2021bxk} further showed that the PDG branching ratios for $K_1(1270)\!\to\!\pi K_0^*(1430)$~\cite{ParticleDataGroup:2024cfk} are inconsistent with a single-state interpretation, reinforcing the two-pole picture.

Together, these theoretical and phenomenological studies establish a coherent framework in which the $K_1(1270)$ resonance emerges as two dynamically generated poles arising from vector-pseudoscalar interactions---an analogue of the well-known two-pole nature of the $\Lambda(1405)$~\cite{Oller:2000fj,Jido:2003cb,Meissner:2020khl,Xie:2023cej,Xie:2023jve,Xie:2025nnq}. 
Despite these advances, direct experimental confirmation remains elusive, and the dependence of the pole structure on light-quark masses---crucial for connecting lattice QCD simulations to experiments---has yet to be systematically explored~\cite{Briceno:2017max,BaryonScatteringBaSc:2023zvt,BaryonScatteringBaSc:2023ori,Zhuang:2024udv}.

Meanwhile, the two-particle femtoscopic correlation technique has become a quantitative and model-sensitive probe of hadron-hadron interactions at low relative momenta~\cite{Kamiya:2019uiw,Liu:2022nec,Vidana:2023olz,Ikeno:2023ojl,Torres-Rincon:2023qll,Liu:2023uly,Liu:2023wfo,Molina:2023oeu,Albaladejo:2024lam,Sarti:2023wlg,Liu:2024uxn,Liu:2024nac,Feijoo:2024bvn,Encarnacion:2024jge,Liu:2025rci,Ge:2025put,Liu:2025oar,Ramos:2025ibe,Shen:2025qpj}. 
Femtoscopy provides direct access to scattering lengths, effective ranges, and near-threshold resonance structures that are otherwise unreachable in conventional scattering experiments. 
Kamiya et al.~\cite{Kamiya:2019uiw} achieved a major step forward by developing a coupled-channel femtoscopic framework for the $K^-\!p$ system considering the $\bar{K}N\!-\!\pi\Sigma\!-\!\pi\Lambda$ dynamics based on chiral SU(3) theory. 
Their results, including Coulomb and coupled-channel effects, successfully reproduced the ALICE $K^-\!p$ correlation data, providing stringent empirical constraints on the $\bar{K}N$ interaction responsible for the $\Lambda(1405)$. 
Following this, Liu et al.~\cite{Liu:2023uly} extended the approach to the $DK$ system, showing that theoretically predicted correlation functions~(CFs) are consistent with interpreting the $D_{s0}^*(2317)$ as a hadronic molecule dynamically generated from the $DK$ interaction, and highlighting that future femtoscopic measurements could directly test this scenario.

Given these developments, applying the femtoscopic framework to the VP interactions offers a natural and timely opportunity to probe the exotic resonance $K_1(1270)$. 
Since direct $\rho K$ and $K^*\pi$ scattering measurements are experimentally inaccessible due to the short lifetimes of vector mesons, information on $K_1(1270)$ must be inferred from the $K\pi\pi$ invariant-mass spectra~\cite{Geng:2006yb,Wang:2019mph,Wang:2020pyy}, where the coupled VP subsystems act as intermediate states. 
In contrast, femtoscopy provides a direct means of accessing the underlying two-body interaction kernels via near-threshold correlation-function observables. 
A systematic study of VP femtoscopic CFs across different channels can thus provide crucial constraints on the two-pole dynamics of $K_1(1270)$ and yield new insights into its internal structure and chiral origin.

In this work, we present a quantitative study of the $K_1(1270)$ resonance within the two-pole scenario, combining chiral extrapolation and femtoscopic CF analyses. 
First, we employ UChPT to trace the mass evolution of the two poles as a function of the pion mass, linking theoretical predictions to future lattice-QCD simulations. 
Second, we analyze femtoscopic CFs of VP systems to probe near-threshold interactions and assess the impact of the two poles on observable correlation functions.

The paper is organized as follows. 
Sec.~\ref{subsec:UChPT} outlines the theoretical framework of VP interactions in UChPT and presents the extrapolation formulas for vector-meson masses. 
Sec.~\ref{subsec:femto} introduces the formalism of femtoscopic CFs and their application to the VP system. 
In Sec.~\ref{subsec:fits}, we fit the chiral trajectories of the vector mesons $\rho$ and $K^*$ using lattice-QCD and experimental data, followed by an analysis of the two-pole evolution with increasing pion mass in Sec.~\ref{subsec:trajs}. 
Sec.~\ref{subsec:femto_num} presents the computed femtoscopic CFs for various coupled channels. 
Finally, Sec.~\ref{sec:summary} summarizes our findings and outlines prospects for future theoretical and experimental studies.

\section{Formalism}
\subsection{UChPT and mass evolution}
\label{subsec:UChPT}
For the vector-pseudoscalar~(VP) interaction, the leading-order~(LO) chiral Lagrangian~\cite{Birse:1996hd,Roca:2005nm} takes the following form
\begin{equation}
    \mathcal{L}_{\mathrm{VP}}^{\mathrm{WT}}
    = -\frac{1}{4f^2}\,
    \mathrm{Tr}\!\left(
        [\mathcal{V}^{\mu},\partial^{\nu}\mathcal{V}_{\mu}]
        [\Phi,\partial_{\nu}\Phi]
    \right),
\end{equation}
from which one obtains the Weinberg-Tomozawa (WT) interaction kernel, projected onto the $S$~wave, which reads~\cite{Roca:2005nm,Geng:2006yb}
\begin{equation}
\begin{aligned}
    V_{ij}(s)
    &= -\,\epsilon^i \!\cdot\! \epsilon^j\,
       \frac{C_{ij}}{8f^2}
       \Bigg[
          3s
          - (M_i^2 + m_i^2 + M_j^2 + m_j^2)  \\
    &\quad
          - \frac{1}{s}\,
            (M_i^2 - m_i^2)(M_j^2 - m_j^2)
       \Bigg],
    \label{eq:VVP}
\end{aligned}
\end{equation}
where $C_{ij}$ are the channel-dependent coupling coefficients determined by SU(3) symmetry~\cite{Roca:2005nm,Geng:2006yb}, $f$ is the decay constant, $M$ and $m$ are the masses of the vector and pseudoscalar mesons, respectively, and $\epsilon^i$ ($\epsilon^j$) denotes the polarization vector of the initial~(final) vector meson.

Within the unitarized chiral perturbation theory~(UChPT)~\cite{Oller:2000ma}, the on-shell unitarized scattering amplitude is given by
\begin{equation}
    T = \big(1 - V G\big)^{-1} V,
    \label{eq:amp}
\end{equation}
where $G$ is a diagonal matrix whose elements $G_i(\sqrt{s})$ correspond to the loop functions of the individual coupled channels.  
In this work, the logarithmic divergence of $G_i(\sqrt{s})$ is regularized using the dimensional-regularization~(DR) scheme, which ensures a proper analytic continuation to higher energies.  
Further details of this formalism can be found in Ref.~\cite{Oller:2000ma}.

An important refinement arises when dealing with intermediate states containing unstable particles with sizable decay widths.  
In such cases, the standard loop functions need to be modified to account for the finite-width effects, 
which can be consistently incorporated through the K\"all\'en--Lehmann spectral representation 
of the vector-meson propagators as
\begin{align}
\tilde{G}_i(\sqrt{s})
&= \frac{1}{C}
\int_{(M_V - 2\Gamma_V)^2}^{(M_V + 2\Gamma_V)^2}
    ds_V \,
    G_i(\sqrt{s}, \sqrt{s_V}, m_i) \notag \\
&\quad \times 
    \left(-\frac{1}{\pi}\right)
    \mathrm{Im}\!\left\{
        \frac{1}{s_V - M_V^2 + i M_V \Gamma_V}
    \right\},
\label{eq:Gi_conv}
\end{align}
where the normalization factor $C$ is given by
\begin{equation}
C =
\int_{(M_V - 2\Gamma_V)^2}^{(M_V + 2\Gamma_V)^2}
    ds_V \,
    \left(-\frac{1}{\pi}\right)
    \mathrm{Im}\!\left\{
        \frac{1}{s_V - M_V^2 + i M_V \Gamma_V}
    \right\}.
\label{eq:C_norm}
\end{equation}
This treatment effectively replaces the narrow-width approximation and leads to a more realistic description of the VP dynamics, as discussed in Refs.~\cite{Roca:2005nm,Geng:2006yb}.

We now turn to the chiral extrapolation of the masses of the relevant particles involved in the $\rho K$ and $K^*\pi$ coupled-channel system.  
As the pion mass increases, the kaon mass approximately follows a linear relation in terms of $m_\pi^2$, which can be expressed as
\begin{equation}
m_K^2 = a + b\,m_\pi^2 + c\,a_{\mathrm{lat}}^2,
\label{eq:mk_fit}
\end{equation}
where the coefficients $a$, $b$, and $c$ are determined by fitting to the lattice-QCD and experimental data.
This form is consistent with the standard chiral extrapolation of pseudoscalar meson masses~\cite{Ren:2012aj}.

For the vector mesons, both the $\rho$ and $K^*$ masses, as well as their couplings relevant to the dominant decay channels $\rho\!\to\!\pi\pi$ and $K^*\!\to\!\pi K$, are parameterized in a similar quadratic form~\cite{Yu:2023xxf,Wang:2025hew,Feng:2010es},  
\begin{equation}
\begin{aligned}
m_{\mathrm{V}} &= c_0 + c_1\,m_\pi^2 + c_2\,a_{\mathrm{lat}}^2, \\[3pt]
g_{\mathrm{VPP}} &= \tilde{c}_0 + \tilde{c}_1\,m_\pi^2 + \tilde{c}_2\,a_{\mathrm{lat}}^2,
\end{aligned}
\label{eq:rho_fit}
\end{equation}
where the coefficients $(c_i,\tilde{c}_i)$ are obtained from global fits combining lattice-QCD results and experimental data. 
Furthermore, the decay width for the $\rho \!\to\!\pi\pi$ decay channel can be written as
\begin{equation}
\Gamma_{\rho} =
\frac{g_{\rho\pi\pi}^2}{6\pi m_{\rho}^2}
\left( \frac{m_{\rho}^2}{4} - m_{\pi}^2 \right)^{3/2},
\end{equation}
and an analogous expression holds for the $K^{*}\!\to\!\pi K$ decay.
In particular, our treatment in Eq.~(\ref{eq:rho_fit}) reproduces the IR-regularized chiral EFT result under the same truncation-keeping terms up to $\mathcal{O}(m_{\pi}^2)$ and discarding higher-order/non-analytic pieces~\cite{Bruns:2004tj}.
Finally, the inclusion of the $a_{\mathrm{lat}}^2$ term effectively accounts for finite-lattice-spacing artifacts up to $\mathcal{O}(a_{\mathrm{lat}}^2)$ and ensures a smooth extrapolation to the continuum limit~\cite{Wang:2025hew}.  

Note that the $\rho$ and $K^*$ mesons can be dynamically generated within the chiral unitary approach or the inverse-amplitude method~(IAM) by employing the chiral Lagrangians for meson-meson scattering, as demonstrated in Refs.~\cite{Oller:1998hw,GomezNicola:2001as}. 
The pion-mass dependence of the $\rho$ meson has been investigated in this framework in Refs.~\cite{Hanhart:2008mx,Guo:2016zos,Molina:2020qpw}. 
In the present work, we focus on tracing the movement of the pole dynamically generated by the WT term of Eq.~(\ref{eq:VVP}) 
in the VP meson interaction. 
For this purpose, a simplified parametrization of the pion-mass dependence of the $\rho$ and $K^*$ mesons is sufficient.

\subsection{Femtoscopic correlation functions}
\label{subsec:femto}

The femtoscopic CF provides a powerful means of extracting information on hadron-hadron interactions at low relative momenta from two-particle momentum correlations.  
Within the standard Koonin--Pratt formalism~\cite{Koonin:1977fh,Pratt:1990zq}, 
the CF between two particles with a relative momentum $\vec{p}$ can be expressed as
\begin{equation}
C(\vec{p}) = \int d^3r\, S(\vec{r}) \, \left| \Psi^{(-)}(\vec{p},\vec{r}) \right|^2,
\label{eq:correlation}
\end{equation}
where $S(\vec{r})$ denotes the source distribution of the relative separation between the two emitted particles at chemical freeze-out, and $\Psi^{(-)}(\vec{p},\vec{r})$ is the outgoing two-body wave function that encodes the final-state interaction~(FSI).  
Assuming a spherically symmetric Gaussian source,
\begin{equation}
S(r) = \frac{1}{(4\pi R^2)^{3/2}} \exp\!\left(-\frac{r^2}{4R^2}\right),
\end{equation}
where $R$ characterizes the source size, the CF depends only on the modulus of the relative momentum $p = |\vec{p}|$.

For coupled-channel systems, such as the VP meson pairs considered here, 
Eq.~(\ref{eq:correlation}) generalizes to~\cite{Albaladejo:2024lam}
\begin{equation}
C_i(p) = \sum_{j} w_j \int d^3r\, S_j(r) 
 \left| \Psi^{(-)}_{ji}(p,r) \right|^2,
\label{eq:coupledC}
\end{equation}
where the indices $i$ and $j$ label the detection and coupled intermediate channels, respectively.  
The outgoing wave function $\Psi^{(-)}_{ji}$ satisfies the Lippmann--Schwinger (Bethe--Salpeter) equation with an interaction potential $V_{ji}$ derived from the on-shell unitarized $T$-matrix in the chiral framework, ensuring consistency between the femtoscopic analysis and the scattering amplitudes that generate the $K_1(1270)$ poles.  
The explicit form of $\Psi^{(-)}_{ji}$ reads
\begin{equation}
\begin{aligned}
\Psi^{(-)}_{ji}(p,r) 
&= \delta_{ji}\, j_0(pr) \\[2pt]
&\quad + \int_0^{q_\mathrm{max}} \!\! \frac{d^3q}{(2\pi)^3}\,
   \frac{E(q)+\omega(q)}
        {E(q)\,\omega(q)} \\[2pt]
&\qquad \times
   \frac{T_{ji}(\sqrt{s},q,p)}
        {s(p)-[E(q)+\omega(q)]^2+i\epsilon}\,
   j_0(qr).
\end{aligned}
\label{eq:psi_short}
\end{equation}
Here, $j_0$ denotes the spherical Bessel function of the first kind.  
The quantities $E(q)=\sqrt{q^2+M_j^2}$ and $\omega(q)=\sqrt{q^2+m_j^2}$ represent the energies of the heavy matter particle (the vector meson) and the pseudoscalar meson in the initial-state channel $j$, respectively.  
The function $s(p)$ corresponds to the squared center-of-mass energy of the final-state channel $i$ with relative momentum $p$.  
$q_\mathrm{max}$ represents a sharp ultraviolet cutoff that suppresses the high-energy components of the propagator.
% Finally, $T_{ji}(\sqrt{s},q,p)$ is the partial-wave scattering amplitude obtained from the unitarized chiral interaction introduced in Eq.~(\ref{eq:amp}) and the Coulomb interaction.

\begin{figure*}[htpb]
    \centering
    \includegraphics[width=3.54in]{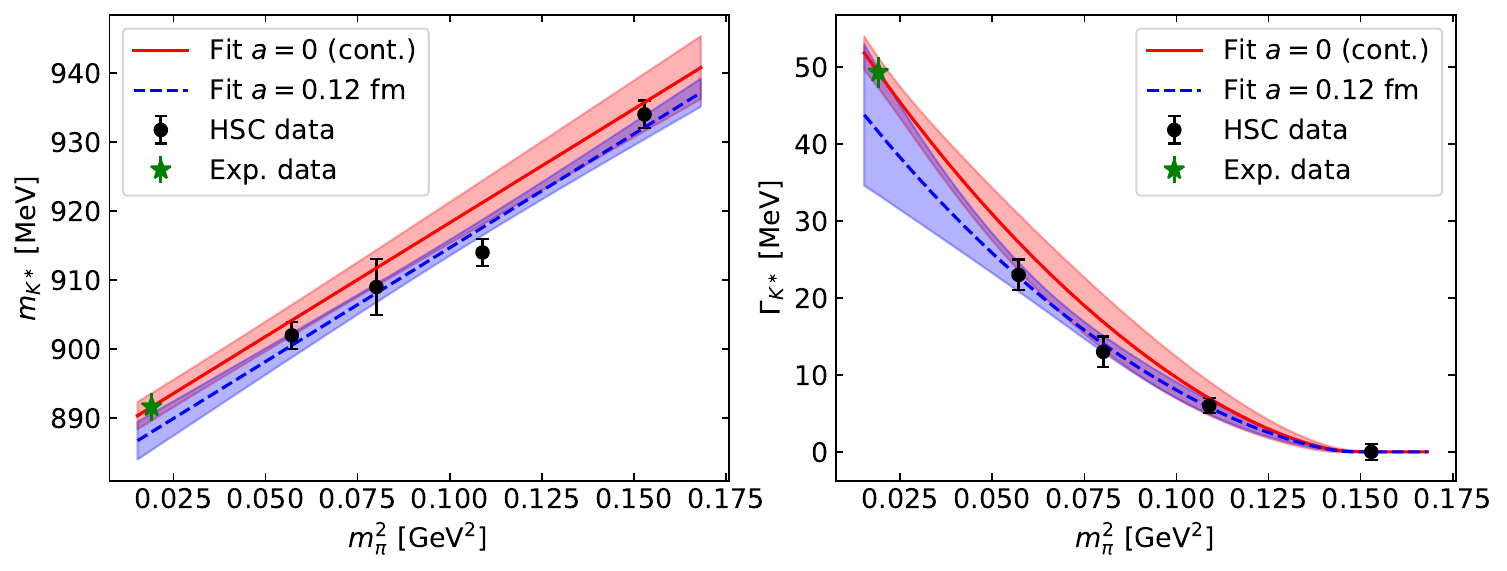}
    \put(-130,-10){(a)}
    \put(121,-10){(b)}
    \hspace{-0.58em}      \includegraphics[width=3.54in]{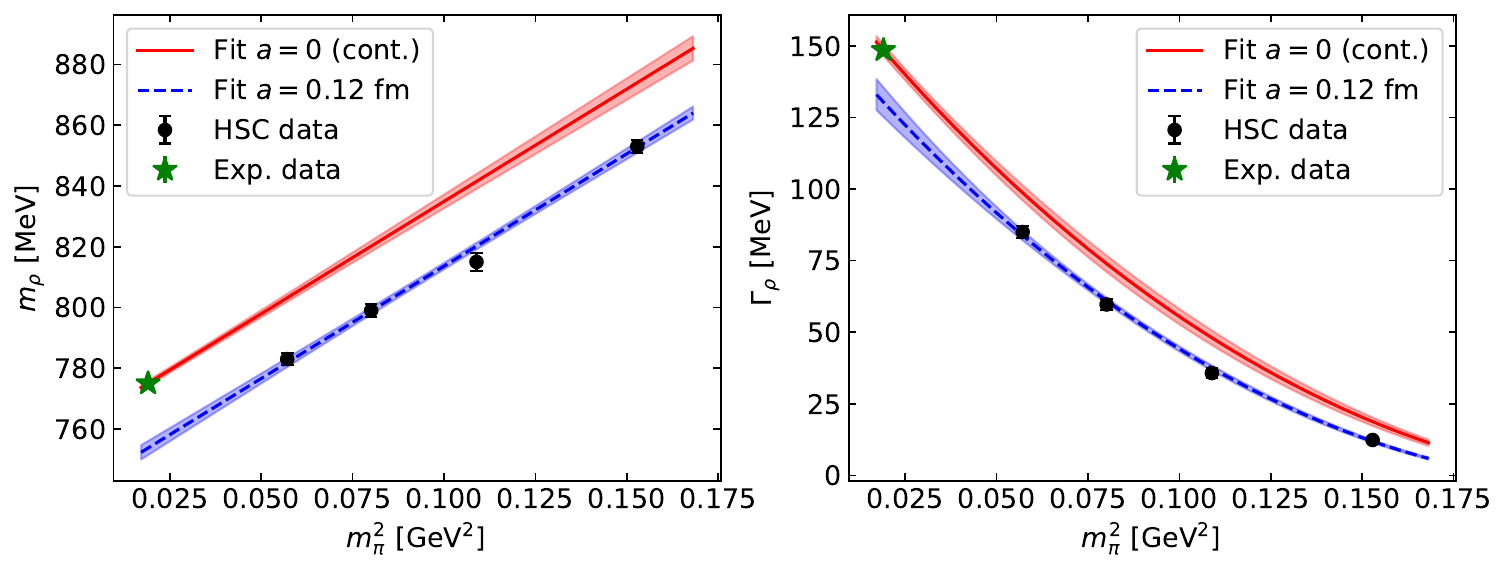}
    \captionsetup{justification=raggedright,singlelinecheck=false}
    \caption{(a) Chiral extrapolation of the $K^{*}$ mass and decay width; (b) chiral extrapolation of the $\rho$ mass and decay width. 
The solid and dashed lines denote the continuum-limit and the discrete lattice QCD fits, respectively, and the shaded regions correspond to the $1\sigma$ uncertainty bands.}
    \label{fig:fit_kstar_rho}
\end{figure*}

In the present work, the strong interaction between the PV pairs is fully incorporated through the coupled-channel $T$-matrix $T^{\mathrm{S}}_{ji}(\sqrt{s})$ of the unitarized chiral perturbation theory in Eq.~(\ref{eq:amp}), while the Coulomb interaction $T^{\mathrm{C}}_{ii}(\sqrt{s},q,p)$ in charged channels (e.g., $\rho^+K^-$ and $K^{*-}\pi^+$) is included perturbatively $T_{ii}(\sqrt{s},q,p) = T^{\mathrm{S}}_{ii}(\sqrt{s}) + T^{\mathrm{C}}_{ii}(\sqrt{s},q,p)$ as described in Refs.~\cite{Torres-Rincon:2023qll,Encarnacion:2024jge}.

\section{Results and Discussion}
\subsection{Fits of vector meson masses}
\label{subsec:fits}

\begin{table}[t]
\centering
\caption{
The tabulated quantities correspond to the pion mass $m_\pi$, kaon mass $m_K$, 
and the extracted $\rho$ and $K^*$ vector-meson pole parameters~(all quantities in units of $\mathrm{MeV}$) 
obtained from the Hadron Spectrum Collaboration~(HSC) studies~\cite{Dudek:2012xn,Wilson:2014cna,Wilson:2015dqa,Wilson:2019wfr,Rodas:2023gma}.
}
\label{tab:lattice_inputs}
\setlength{\tabcolsep}{5.3pt}
\renewcommand{\arraystretch}{1.15}
\begin{tabular}{cccccc}
\hline\hline
$m_\pi$ & $m_K$ & $m_\rho$ & $\Gamma_\rho$ & $m_{K^{*}}$ & $\Gamma_{K^{*}}$ \\
\hline
239 & $507.7 \pm 2.4$ & $783 \pm 2$ & $85.0 \pm 2.0$ & $902 \pm 2$ & $23 \pm 2$ \\
283 & $519.4 \pm 2.6$ & $799 \pm 2$ & $59.7 \pm 1.9$ & $909 \pm 4$ & $13 \pm 2$ \\
330 & $527.8 \pm 2.8$ & $815 \pm 3$ & $35.8 \pm 1.8$ & $914 \pm 2$ & $6 \pm 1$ \\
391 & $549.1 \pm 1.2$ & $853 \pm 2$ & $12.4 \pm 0.6$ & $934 \pm 2$ & $0 \pm 1$ \\
\hline\hline
\end{tabular}
\end{table}

According to the $\pi\pi$ and $\pi K$ scattering phase shifts obtained from lattice QCD simulations performed by the Hadron Spectrum Collaboration~(HSC) over the past decade~\cite{Dudek:2012xn,Wilson:2014cna,Wilson:2015dqa,Wilson:2019wfr,Rodas:2023gma}, 
the masses of the $\rho$ and $K^*$ vector mesons can be extracted from the pole positions of parametrized scattering amplitudes 
evaluated at different pion and kaon masses.

Based on the results summarized in Table~\ref{tab:lattice_inputs}, 
we carry out a global fit that combines the HSC lattice QCD results with the corresponding experimental measurements, 
following the strategy outlined in Sec.~\ref{subsec:UChPT}~\footnote{Note that the HSC does not provide scattering data at different lattice spacings, and therefore this assumption in Eq.~(\ref{eq:rho_fit}) should be verified in future lattice QCD simulations.}. 
In this fit, the kaon trajectory is described by the linear squared–mass relation in Eq.~(\ref{eq:mk_fit}), 
whereas the $\rho$ and $K^*$  are modeled by the quadratic forms in Eq.~(\ref{eq:rho_fit}) for their masses and couplings, 
which effectively capture the impact of their sizable decay widths. 
Since the HSC simulations were performed at a finite spatial lattice spacing $a_{\mathrm{lat}}\sim 0.12
~\mathrm{fm}$~(temporal spacing $a_t$ is used for the scale setting)~\cite{Dudek:2012xn,Wilson:2014cna,Wilson:2015dqa,Wilson:2019wfr,Rodas:2023gma}, 
we include an $a_{\mathrm{lat}}^{2}$ term to account for discretization corrections and to provide continuum limit inputs to the UChPT framework. 
The fit parameters obtained from this global analysis are listed in Table~\ref{tab:mass_fits}, 
and the corresponding curves with the 1$\sigma$ uncertainty bands are shown in Fig.~\ref{fig:fit_kstar_rho}. 
The resulting masses and couplings as functions of $m_\pi$ are then used as input to the coupled-channel UChPT study in the next section.

\begin{table}[t]
\centering
% \captionsetup{justification=raggedright,singlelinecheck=false}
\caption{Fit coefficients for the $m_\pi$-dependence of pseudoscalar and vector meson masses and couplings 
in the continuum limit (all quantities in units of $\mathrm{GeV}$).}
\label{tab:mass_fits} 
\setlength{\tabcolsep}{6pt}
\renewcommand{\arraystretch}{1.2}
\begin{tabular}{lccc}
\hline\hline
Particle & Parameter & Value $\pm$ Uncertainty & Unit \\
\hline
$K$ 
& $a_K$ & $0.23697 \pm 0.00048$ & $\mathrm{GeV}^{2}$ \\
& $b_K$ & $0.454 \pm 0.025$ & 1 \\
& $c_K$ & $-0.0137 \pm 0.0075$ & $\mathrm{GeV}^{4}$ \\
\hline
$\rho$ 
& $c^{(\rho)}_{0}$ & $0.7609 \pm 0.0010$ & $\mathrm{GeV}$ \\
& $c^{(\rho)}_{1}$ & $0.740 \pm 0.027$ & $\mathrm{GeV}^{-1}$ \\
& $c^{\rho)}_{2}$ & $-0.0576 \pm 0.0070$ & $\mathrm{GeV}^{3}$ \\
& $\tilde{c}^{(\rho)}_{0}$ & $5.939 \pm 0.054$ & 1 \\
& $\tilde{c}^{(\rho)}_{1}$ & $0.60 \pm 1.8$ & $\mathrm{GeV}^{-2}$ \\
& $\tilde{c}^{(\rho)}_{2}$ & $-0.71 \pm 0.34$ & $\mathrm{GeV}^{2}$ \\
\hline
$K^{*}$ 
& $c^{(K^{*})}_{0}$ & $0.8854 \pm 0.0021$ & $\mathrm{GeV}$ \\
& $c^{(K^{*})}_{1}$ & $0.330 \pm 0.029$ & $\mathrm{GeV}^{-1}$ \\
& $c^{(K^{*})}_{2}$ & $-0.0101 \pm 0.0091$ & $\mathrm{GeV}^{3}$ \\
& $\tilde{c}^{(K^{*})}_{0}$ & $5.61 \pm 0.21$ & 1 \\
& $\tilde{c}^{(K^{*})}_{1}$ & $-1.6 \pm 9.1$ & $\mathrm{GeV}^{-2}$ \\
& $\tilde{c}^{(K^{*})}_{2}$ & $-1.4 \pm 1.4$ & $\mathrm{GeV}^{2}$ \\
\hline\hline
\end{tabular}
\end{table}

% table 1

\begin{figure*}[htpb]
    \centering
    \includegraphics[width=3.4in]{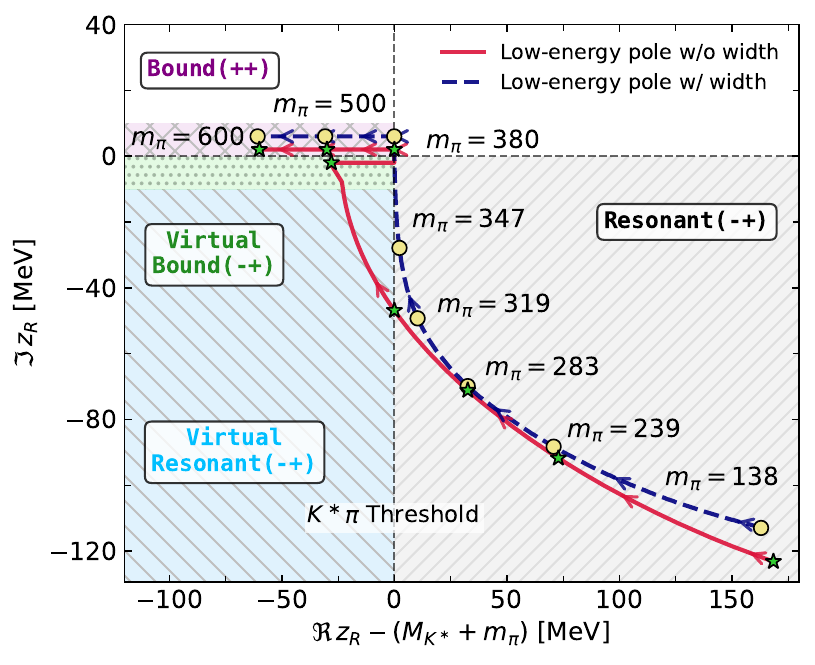}\quad 
       \includegraphics[width=3.4in]{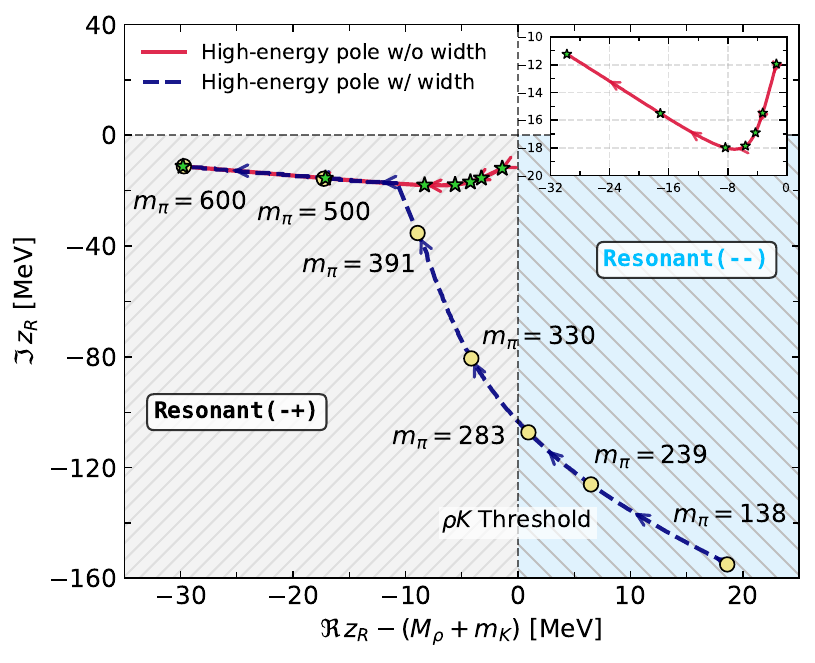}
       % \captionsetup{justification=raggedright,singlelinecheck=false}
    \caption{Chiral evolution of the two-pole structure of the $K_{1}(1270)$ resonance in the $\rho K$--$K^{*}\pi$ coupled-channel system. 
The left and right panels correspond to the lower and higher poles, mainly coupled to $K^{*}\pi$ and $\rho K$, respectively. 
Solid and dashed lines denote the results without and with finite vector-meson widths, respectively. 
Markers indicate pole positions at different pion masses, with the critical masses explicitly labeled. 
}
    \label{fig:trajectory}
\end{figure*}

\subsection{Two-pole trajectories }
\label{subsec:trajs}
In Ref.~\cite{Geng:2006yb}, the coupled-channel analysis of the vector-pseudoscalar~(VP) interaction was performed within the leading-order unitarized chiral perturbation theory.  
Two poles were dynamically generated, located just below the $\rho K$ threshold and slightly above the $K^*\pi$ threshold, respectively.  
When the three higher channels are neglected, the essential features of the spectrum remain essentially unchanged. 
By slightly readjusting the subtraction constants to $a_{K^*\pi} = -2.21$ and $a_{\rho K} = -2.44$ while keeping $\mu = 900~\mathrm{MeV}$ and $f = 115~\mathrm{MeV}$,
one obtains nearly identical pole positions at 
$W_H = 1269.5 - 12.0i~\mathrm{MeV}$ and $W_L = 1198.5 - 123.2i~\mathrm{MeV}$ 
for the zero-width case~\cite{Xie:2023cej}. 
For comparison, if we keep the same set of subtraction constants as above, 
the pole positions shift to 
$W_H = 1289.2 - 154.0i~\mathrm{MeV}$ and $W_L = 1192.6 - 113.3i~\mathrm{MeV}$ 
in the finite-width case. 
The sizable shift of the higher pole is of kinematical origin and does not affect the underlying two-channel dynamics~\cite{Geng:2006yb}, as will be illustrated by the pole trajectories.
This demonstrates that the dominant dynamics of the $K_1(1270)$ resonance 
are already well captured by the reduced two-channel system composed of $\rho K$ and $K^*\pi$.

Motivated by this observation, we perform the present calculation in the same unitarized chiral framework, 
restricting ourselves to the $\rho K$ and $K^*\pi$ coupled channels.  
The rationale for this simplification is twofold.  
First, as will be demonstrated in Sec.~\ref{subsec:femto} through the study of femtoscopic CFs, 
the dominant contributions to the femtoscopic line shapes arise from these two channels, 
whose distinct line shapes reflect their different coupling strengths~\cite{Xie:2023cej}.  
Second, from the lattice-QCD perspective, simulating multi-channel scattering involving unstable vector mesons such as $\rho$ and $K^*$ is extremely challenging, 
and current simulations are limited to systems composed of stable hadrons at relatively low pion masses~\cite{BaryonScatteringBaSc:2023zvt,BaryonScatteringBaSc:2023ori}.
Therefore, the two-channel approximation provides both a physically motivated and practically tractable description of the $K_1(1270)$ dynamics.

Furthermore, the essential difference between the $K_1(1270)$ system and the other well-known two-pole structure, the $\Lambda(1405)$~\cite{Xie:2023cej}, 
lies in the fact that the heavy scattering constituents in the former case, $\rho$ and $K^*$, are themselves resonances with large decay widths and short lifetimes.  
To clarify the chiral trajectories of the lower and higher poles, we perform calculations in two scenarios: 
one neglecting and one incorporating the vector meson decay widths.  
We expect that the qualitative evolution behavior of the poles is largely independent of whether or not the vector-meson widths are explicitly included. Fortunately, the unitarized chiral framework provides a straightforward way to incorporate the finite widths and spin properties of the $\rho$ and $K^*$ mesons 
by modifying the analytic form of the loop functions as described in Sec.~\ref{subsec:UChPT}.  

In Fig.~\ref{fig:trajectory}, we show the resulting chiral trajectories of the two poles.  
The left panel corresponds to the lower pole, which couples predominantly to the $K^*\pi$ channel,  
while the right panel corresponds to the higher pole, mainly associated with the $\rho K$ channel.  
The evolution of both poles is consistent with that reported in Ref.~\cite{Xie:2023cej} for the $\Lambda(1405)$, highlighting their common origin from chiral dynamics.  
In both subfigures, different pion-mass values along the trajectories are indicated by stars or circles.

\begin{figure*}[htpb]
    \centering
    \includegraphics[width=1.79in]{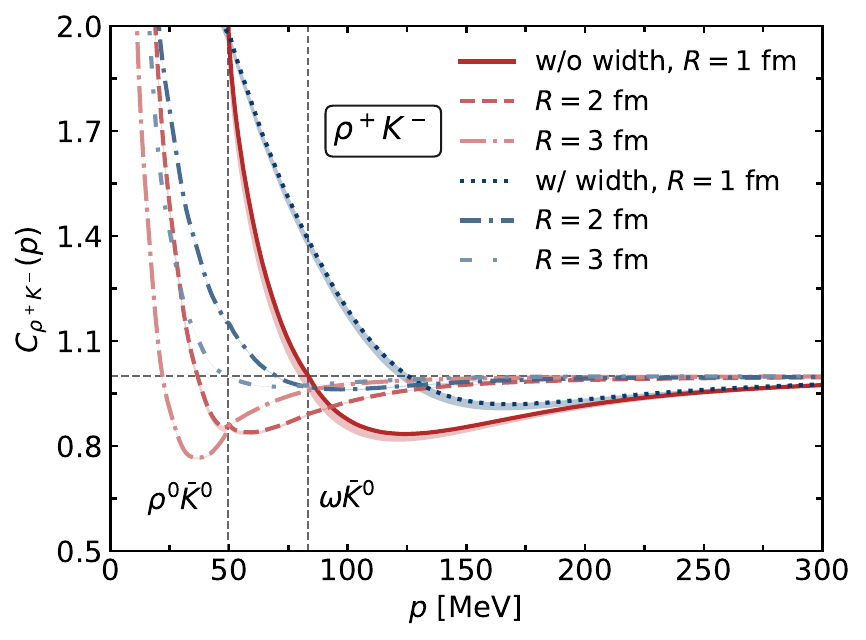}  \hspace{-0.76em}      \includegraphics[width=1.79in]{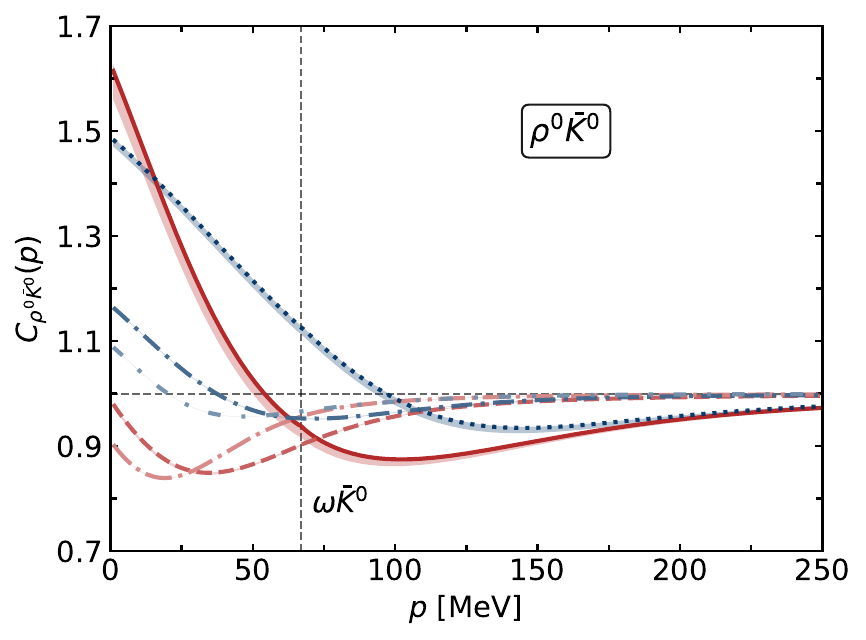} \hspace{-0.76em}
    \includegraphics[width=1.79in]{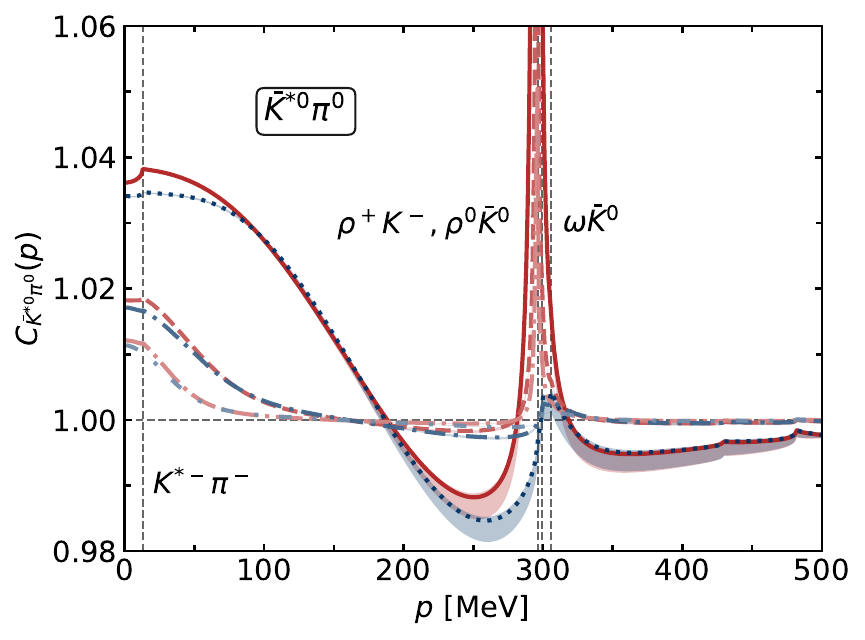}  \hspace{-0.76em}
       \includegraphics[width=1.79in]{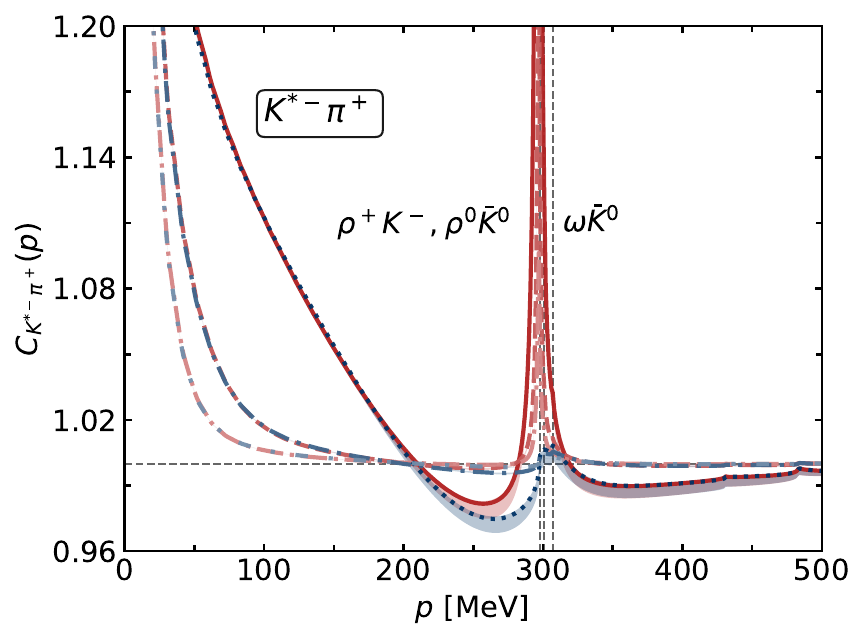} \\
    \includegraphics[width=1.79in]{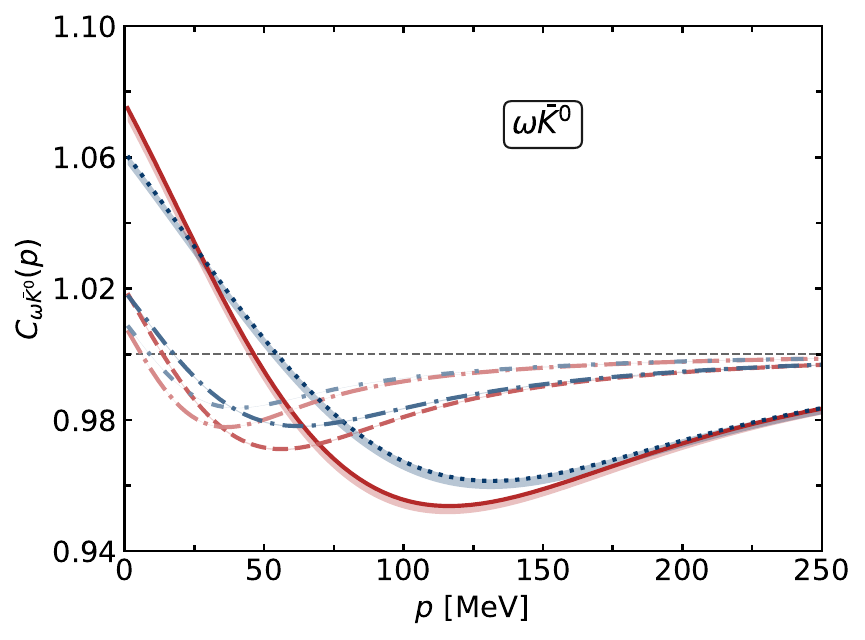}  \hspace{-0.76em}      \includegraphics[width=1.79in]{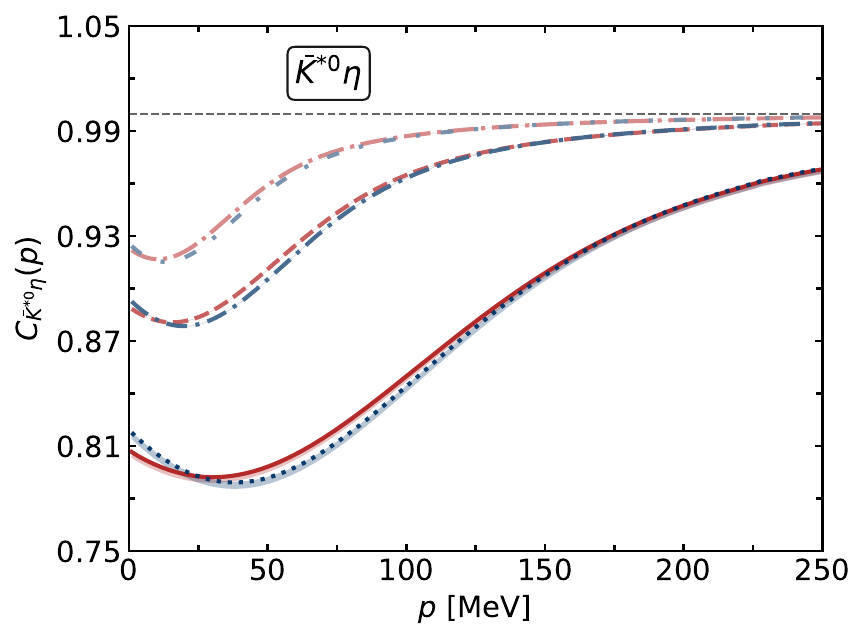} \hspace{-0.76em}
    \includegraphics[width=1.79in]{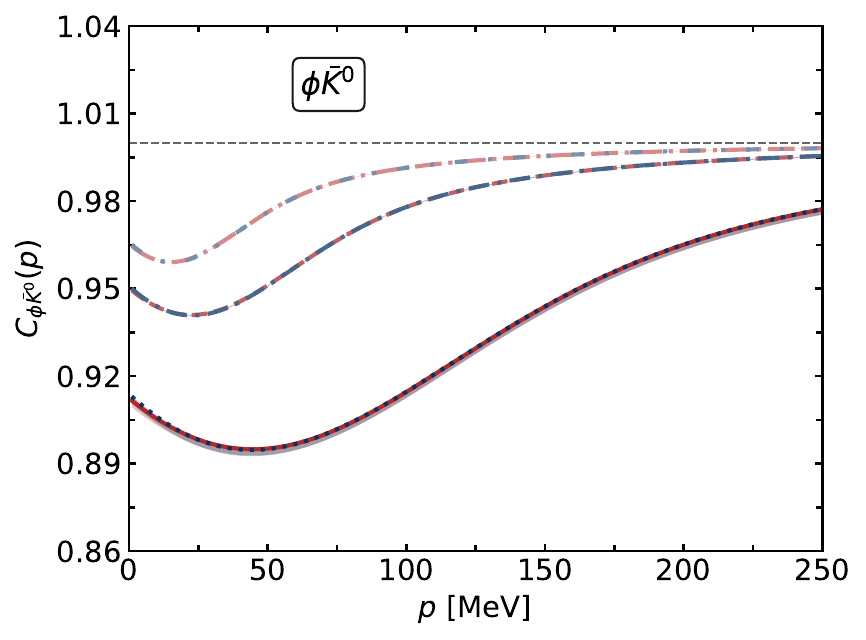}  \hspace{-0.76em}
       \includegraphics[width=1.79in]{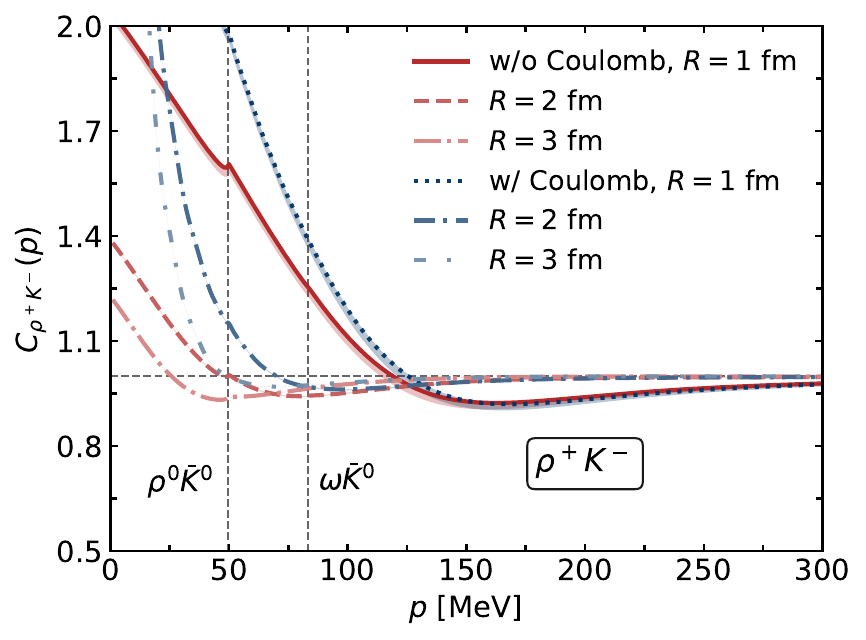} 
       % \captionsetup{justification=raggedright,singlelinecheck=false}
    \caption{Femtoscopic correlation functions $C_i(p)$ for the $Q = 0$ sector, 
    corresponding to the coupled-channel system
    $\{\rho^{+} K^{-},\allowbreak\, \rho^{0}\bar{K}^{0},\allowbreak\, 
\bar{K}^{*0}\pi^{0},\allowbreak\, K^{*-}\pi^{+},\allowbreak\, 
\omega \bar{K}^{0},\allowbreak\, \bar{K}^{*0}\eta,\allowbreak\, 
\phi \bar{K}^{0}\}$. 
    The red and blue curves represent results without and with vector-meson widths, respectively, and different line transparencies correspond to source sizes of $R = 1$, $2$, and $3~\mathrm{fm}$.
    }
    \label{fig:CFQ0}
\end{figure*}

In the left panel, the lower pole exhibits a characteristic behavior:  
as the pion mass increases, the strength of the Weinberg–Tomozawa~(WT) interaction grows,  
causing the pole to move from an above-threshold resonance to a below-threshold virtual or bound state.  
In the case without vector-meson widths, the trajectory undergoes a ``zigzag'' evolution:  
the resonance first becomes a below-threshold state located on the second Riemann sheet ($-+$),  
then evolves into a virtual state as the width decreases to zero,  
and eventually crosses to the first Riemann sheet ($++$) as a zero-energy bound state before becoming a deeply bound state as the binding energy increases~\cite{Xie:2023cej}.  
When the widths are included, the pole instead evolves directly from an above-threshold resonance to a bound state without exhibiting the zigzag pattern,  
in agreement with the general expectation discussed in Ref.~\cite{Hyodo:2013iga}.

The right panel shows a much simpler pattern for the higher pole, consistent with the findings of Ref.~\cite{Xie:2023cej}.  
As the pion mass increases, the attraction in the $\rho K$ channel strengthens, 
and the higher pole moves downward in energy, following a more intuitive trajectory.  
The main difference between the two scenarios lies in the physical starting point:  
in the case without width, the higher pole initially appears as a shallow below-threshold resonance,  
while in the width-included case, it starts as an above-threshold resonance with a larger imaginary part.  
Nevertheless, both treatments yield essentially the same qualitative two-pole trajectories, 
demonstrating that the inclusion of finite widths does not significantly alter the overall chiral evolution of the $K_1(1270)$ poles.

Since the vector-meson mass evolution formulae include uncertainties propagated from the fits to the lattice-QCD data~\cite{Dudek:2012xn,Wilson:2014cna,Wilson:2015dqa,Wilson:2019wfr,Rodas:2023gma}, 
we also take into account the corresponding uncertainties in the pole trajectories.  
However, we do not explicitly display the error bands in the trajectory Fig.~\ref{fig:trajectory}, 
because the variations of the pole positions are relatively small---less than $5~\mathrm{MeV}$ in both mass and width---and thus visually indistinguishable on the scale of the figure.

\subsection{Femtoscopic correlation functions}
\label{subsec:femto_num}
Since the femtoscopic CF is a direct experimental observable, 
it is necessary to convert the theoretical amplitudes from the isospin basis, 
which is more suitable for lattice-QCD studies with an $N_f = 2 + 1$ flavor setup~\cite{Briceno:2017max}, 
to the charge basis, which corresponds to experimentally measurable final states~\cite{Liu:2024uxn}.  
The $K_1(1270)$ forms an isospin doublet with two members, $K_1^0(1270)$~\cite{Wang:2020pyy,Dias:2021upl} and $K_1^{-}(1270)$~\cite{Geng:2006yb,Wang:2019mph}.  
For $I_3\allowbreak =\allowbreak +1/2\,~\allowbreak(Q=0)$, the channels are 
$\{\rho^{+} K^{-},\allowbreak\, \rho^{0}\bar{K}^{0},\allowbreak\, 
\bar{K}^{*0}\pi^{0},\allowbreak\, K^{*-}\pi^{+},\allowbreak\, 
\omega \bar{K}^{0},\allowbreak\, \bar{K}^{*0}\eta,\allowbreak\, 
\phi \bar{K}^{0}\}$,
while for $I_3\allowbreak =\allowbreak -1/2\,~\allowbreak(Q=-1)$ they are 
$\{\rho^{0} K^{-},\allowbreak\, \rho^{-}\bar{K}^{0},\allowbreak\, 
K^{*-}\pi^{0},\allowbreak\, \bar{K}^{*0}\pi^{-},\allowbreak\, 
\omega K^{-},\allowbreak\, K^{*-}\eta,\allowbreak\, 
\phi K^{-}\}$.

To include the finite-width effects of the vector mesons, 
we adopt the formalism developed in Ref.~\cite{Feijoo:2024bvn}, 
which consistently accounts for the widths both in the $T$-matrix and in the asymptotic final-state relative wave functions, 
ultimately yielding modified femtoscopic CFs.  
For the $I_3 = +1/2$ case, in addition to the strong interactions that are common to both isospin components, 
the electromagnetic Coulomb interaction must also be considered due to the presence of the charged channels $\rho^+ K^-$ and $K^{*-}\pi^+$.  
The implementation of the Coulomb contribution follows the prescriptions given in Sec.~\ref{subsec:femto},
where it is incorporated perturbatively by adding the Coulomb amplitudes to the diagonal strong scattering amplitudes in the same manner as a Born-term correction.

\begin{figure*}[htpb]
    \centering
    \includegraphics[width=1.79in]{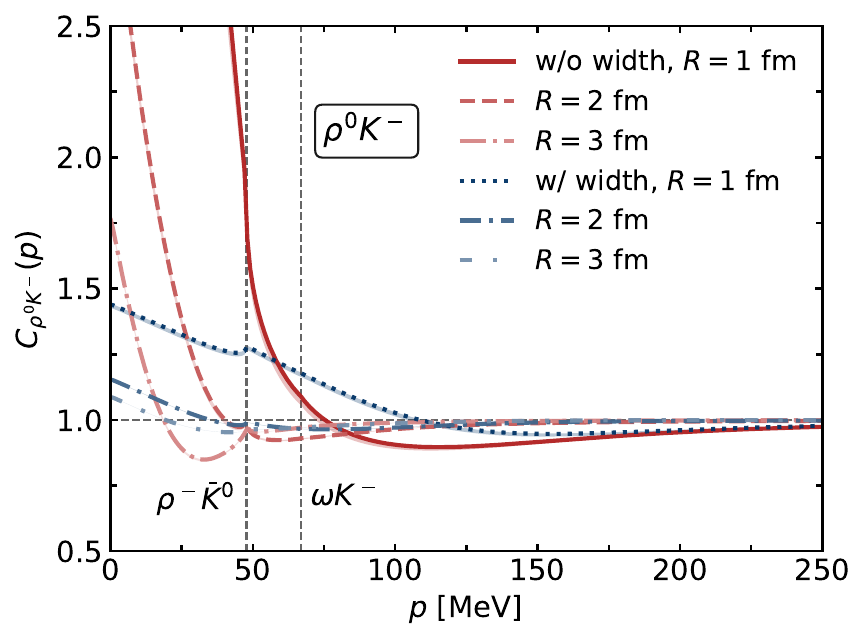}  \hspace{-0.76em}      \includegraphics[width=1.79in]{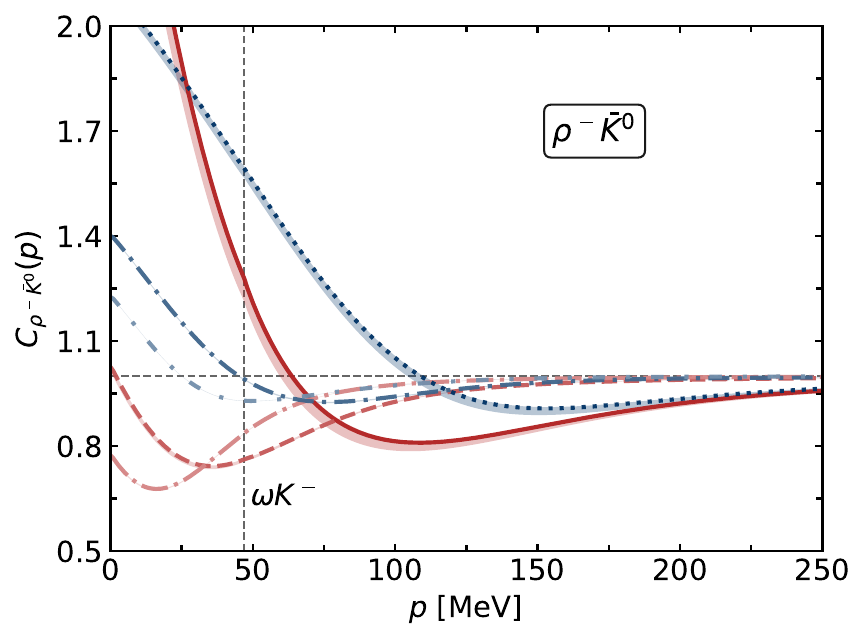} \hspace{-0.76em}
    \includegraphics[width=1.79in]{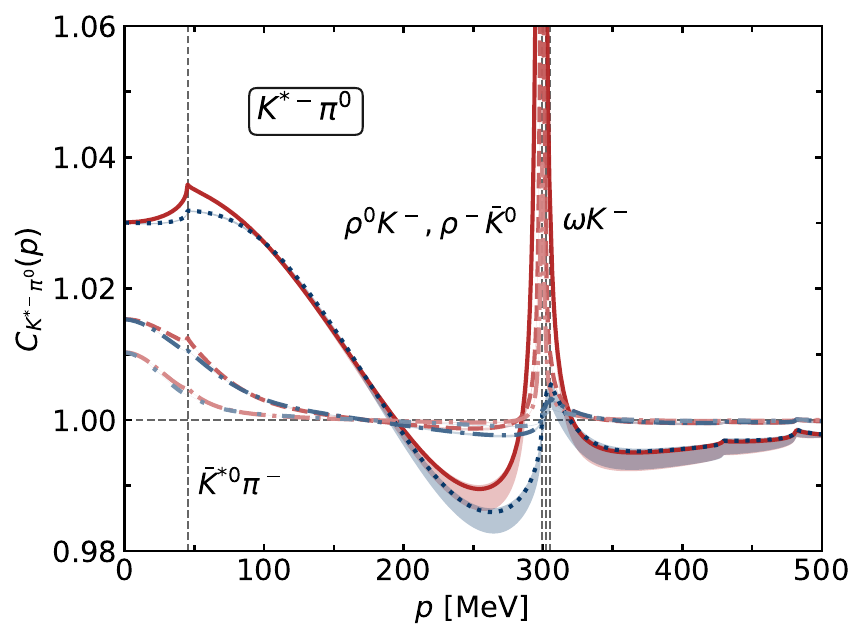}  \hspace{-0.76em}
       \includegraphics[width=1.79in]{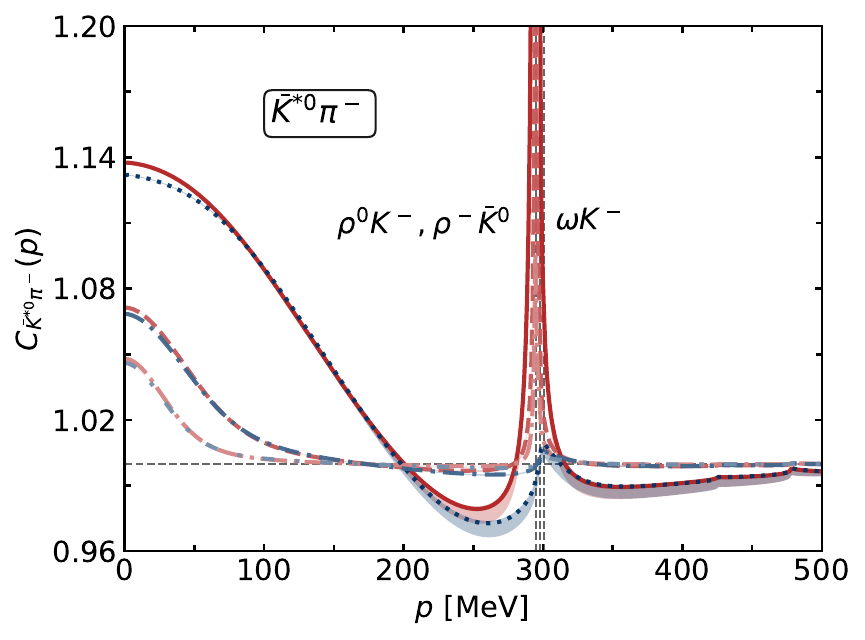} \\
    \includegraphics[width=1.79in]{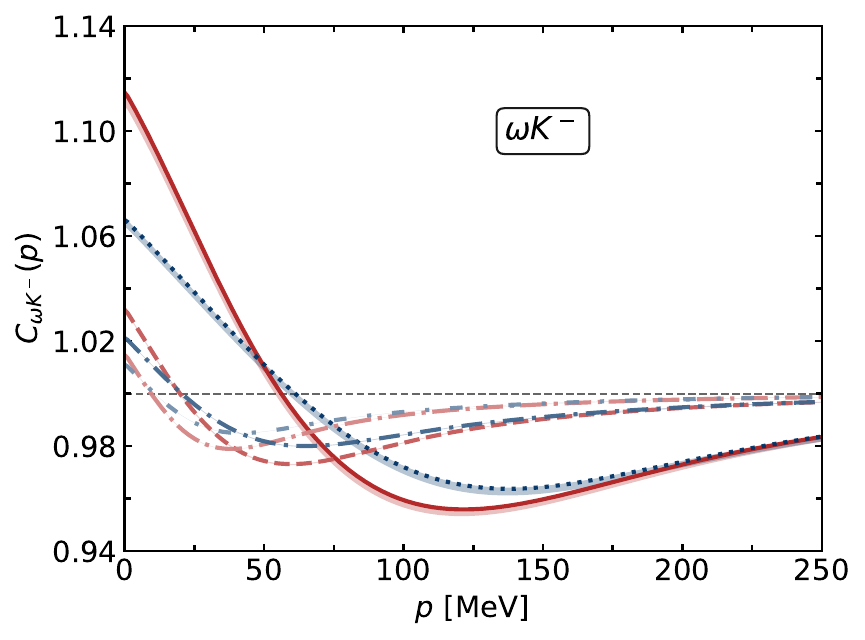}  \hspace{-0.76em}      \includegraphics[width=1.79in]{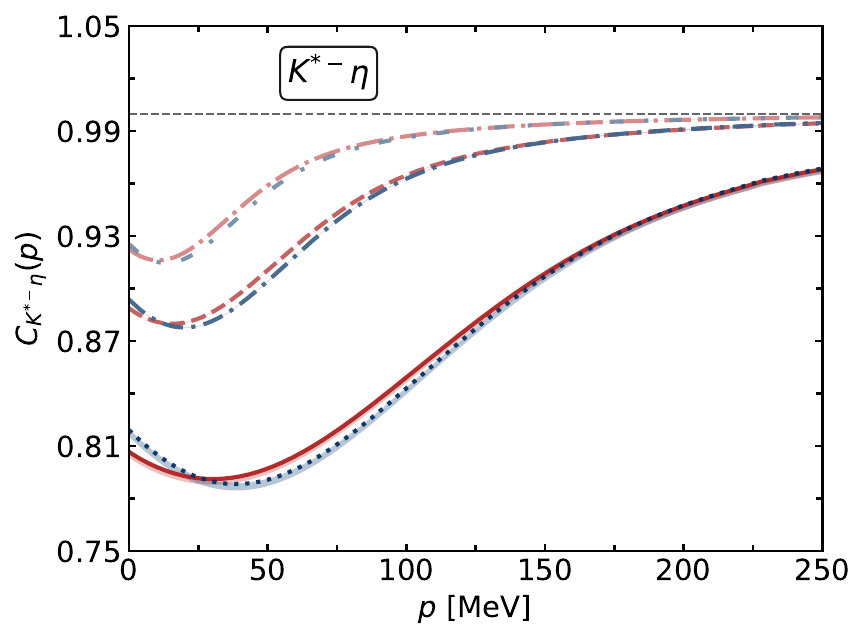} \hspace{-0.76em}
    \includegraphics[width=1.79in]{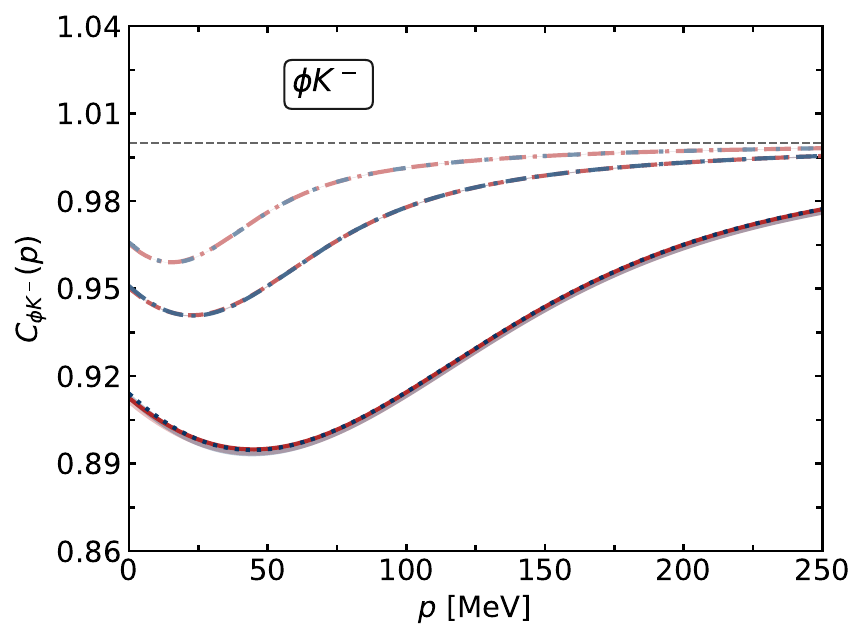}  \hspace{-0.76em}
       \includegraphics[width=1.79in]{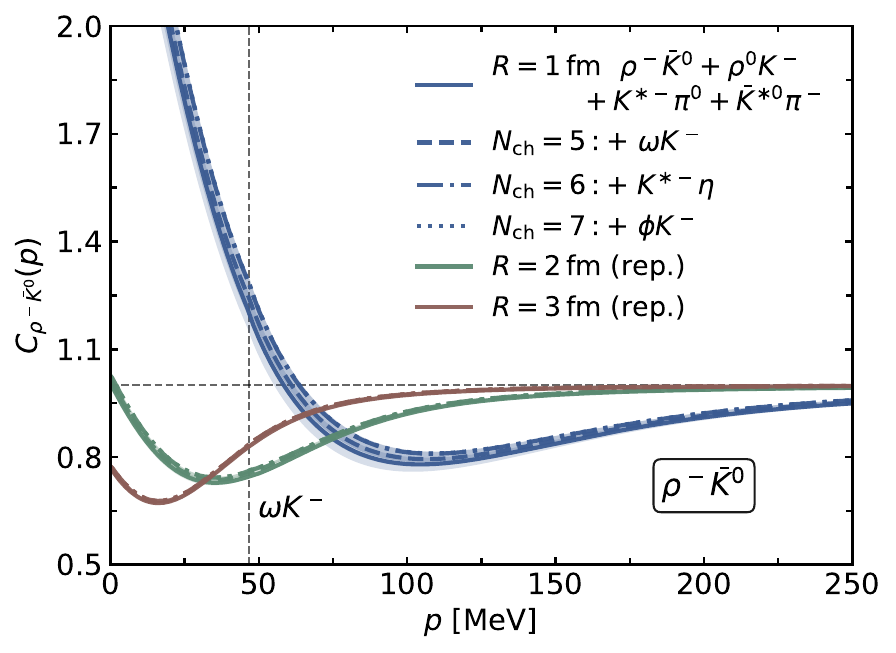} 
       % \captionsetup{justification=raggedright,singlelinecheck=false}
    \caption{Femtoscopic correlation functions $C_i(p)$ for the $Q = -1$ sector, corresponding to the coupled-channel system 
    $\{\rho^{0} K^{-},\allowbreak\, \rho^{-}\bar{K}^{0},\allowbreak\, 
K^{*-}\pi^{0},\allowbreak\, \bar{K}^{*0}\pi^{-},\allowbreak\, 
\omega K^{-},\allowbreak\, K^{*-}\eta,\allowbreak\, 
\phi K^{-}\}$.
    The red and blue curves denote calculations without and with the finite vector-meson widths, respectively, 
    and different line transparencies correspond to source sizes of $R = 1$, $2$, and $3~\mathrm{fm}$.}
    \label{fig:CFQm1}
\end{figure*}

We summarize here the relevant parameters employed in the calculation of the femtoscopic CFs.  
First, the source size $R$ is taken to be three representative values $R = 1, 2, 3~\mathrm{fm}$.
Second, the channel weights $w_j$, which quantify the relative contributions from various coupled channels, are set to unity for simplicity.  
Third, the subtraction constants, renormalization scale, and decay constant are chosen to be consistent with those adopted in Ref.~\cite{Geng:2006yb}, namely
\begin{equation}
\mu = 900~\mathrm{MeV}, \quad
a(\mu) = -1.85, \quad
f = 115~\mathrm{MeV},
\end{equation}
where $\mu$ denotes the renormalization scale, $a(\mu)$ is the universal subtraction constant, and $f$ represents the pion decay constant.  
These parameter choices follow the convention established in previous analyses of the $K_1(1270)$ system~\cite{Wang:2019mph,Wang:2020pyy,Dias:2021upl} and ensure consistency with the chiral interaction framework used throughout this study.
Finally, to estimate the theoretical uncertainties, 
we vary the cutoff $q_\mathrm{max}$ in Eq.~(\ref{eq:psi_short}) within a reasonable range of $0.6$--$1.4~\mathrm{GeV}$. 
This variation is consistent with the typical choice of the subtraction constant $a(\mu)$, 
and corresponds to the equivalent cutoff $\Lambda$ in the cutoff regularization scheme of the $T$-matrix~\cite{Roca:2005nm}.

In Figs.~\ref{fig:CFQ0} and \ref{fig:CFQm1}, we show the calculated CFs for the seven final-state channels in the two charge bases $Q=0$ and $Q=-1$, respectively.  
For each channel, the CF $C_i(p)\ (i = 1,\dots,7)$ is plotted over the relative momentum range $(0,250)~\mathrm{MeV}$, 
except for the $\bar{K}^*\pi$ channels, for which the range $(0,500)~\mathrm{MeV}$ is adopted to illustrate better the threshold cusps induced by the coupled $\rho\bar{K}$ and $\omega\bar{K}$ channels.  
These cusp structures are significantly smoothened when the vector-meson decay widths are included.  
The threshold effects from the higher channels $\bar{K}^*\eta$ and $\phi\bar{K}$ are omitted in all plots, 
as their influence on the line shapes of the CFs is negligible.

In both Figs.~\ref{fig:CFQ0} and \ref{fig:CFQm1}, two sets of curves are shown using distinct color schemes: 
the red curves correspond to calculations without including the vector-meson widths, 
and the blue curves include the width effects.  
Within each color set, different source sizes $R = 1,\, 2,\, 3~\mathrm{fm}$ are plotted with decreasing opacity.  
From the first five subfigures, one can observe that the red curves are much sharper than the blue ones.  
This behavior reflects that incorporating the finite widths of the intermediate vector mesons effectively smears the loop functions, 
resulting in smoother scattering amplitudes.  
The only exceptions are the highest-threshold channels, $\bar{K}^*\eta$ and $\phi\bar{K}$, 
where the couplings to the lower channels are weak, as evident from the potential-coefficient matrix $C_{ij}$ in Eq.~(\ref{eq:VVP}).

Within each color system, increasing the source size $R$ systematically weakens the interaction strength, 
leading to a corresponding reduction in the magnitude of the CF, 
as expected from the standard source-size dependence in Femtoscopy~\cite{Kamiya:2019uiw,Liu:2023wfo}.

Comparing the subfigures for different final-state channels, 
one can see that for the five lower-threshold channels, 
the CFs exhibit resonant-like behavior, 
consistent with the findings of Refs.~\cite{Liu:2023uly,Liu:2024nac}.  
In contrast, for the last two channels, $\bar{K}^*\eta$ and $\phi\bar{K}$, 
the line shapes resemble those of weakly bound states due to the small coupling between the two poles and these higher channels.

Turning to the $Q = 0$ sector in Fig.~\ref{fig:CFQ0}, a pronounced enhancement appears in the low-momentum region of the CFs in channels $\rho^+ K^-$ and $K^{*-}\pi^+$, 
as compared with their isospin-partner channels $\rho^0K^-$ and $\bar{K}^{*0}\pi^-$ in Fig.~\ref{fig:CFQm1}~\cite{Kamiya:2019uiw}.
This enhancement originates predominantly from the Coulomb interaction rather than from small isospin-breaking effects.
To further illustrate this behavior, the last subfigure in Fig.~\ref{fig:CFQ0} shows the $\rho^+K^-$ CF with and without the Coulomb interaction, 
both calculated including the finite vector-meson widths. 
The comparison confirms that the Coulomb force produces a strong near-threshold enhancement, 
whereas the finite widths mainly smooth the overall correlation pattern.

In the last subfigure of the $Q=-1$ sector of Fig.~\ref{fig:CFQm1}, 
three color groups (blue, green, and yellow) represent source sizes of $R = 1,\, 2,\, 3~\mathrm{fm}$, respectively.  
Within each group, we sequentially add channels to the coupled system with finite widths: 
starting with only $\rho\bar{K}$ and $\bar{K}^*\pi$, and including $\omega\bar{K}$, $\bar{K}^*\eta$, and $\phi\bar{K}$~\cite{Kamiya:2019uiw,Liu:2023wfo}, one by one.  
The resulting curves lie almost on top of each other, showing only minor increases in magnitude as additional channels are introduced.  
This confirms that restricting the analysis to the dominant two channels, $\rho\bar{K}$ and $\bar{K}^*\pi$, is a justified and physically sound approximation~\cite{Xie:2023cej}.

\begin{figure}[htpb]
    \centering
    \includegraphics[width=1.71in]{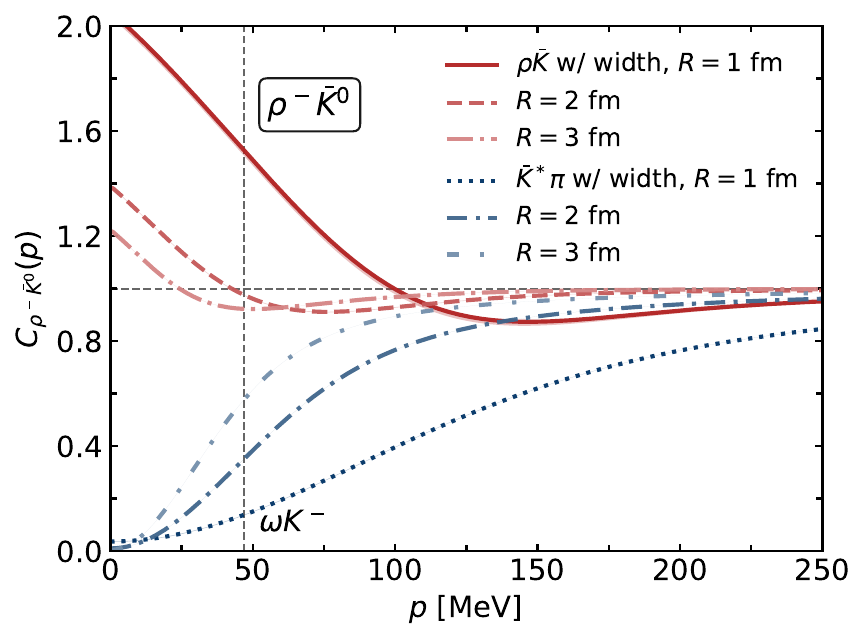}
    \hspace{-0.7em}      \includegraphics[width=1.71in]{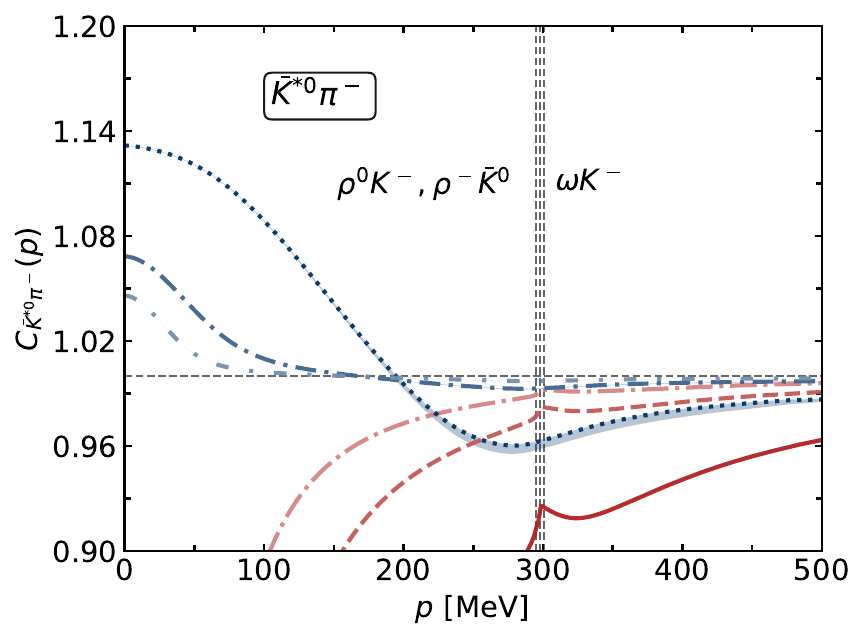}
    \captionsetup{justification=raggedright,singlelinecheck=false}
    \caption{Femtoscopic correlation functions $C_i(p)$ for the two channels 
    $\rho^-\bar{K}^0$ and $\bar{K}^{*0}\pi^-$ in the $Q=-1$ system. 
    The red and blue lines represent the individual contributions to $C(p)$ from the 
    $\rho\bar{K}$ and $\bar{K}^*\pi$ channels, respectively. 
    Different line transparencies correspond to source radii of 
    $R = 1$, $2$, and $3~\mathrm{fm}$.}
    \label{fig:CF_twopole}
\end{figure}

Finally, to clarify more precisely the connection between the two-pole structure and the femtoscopic CFs, we focus on two representative channels, $\rho^-\bar{K}^0$ and $\bar{K}^{*0}\pi^-$, from the $Q=-1$ system, and calculate the individual contributions from the $\rho\bar{K}$ and $\bar{K}^*\pi$ components, respectively. 
From Fig.~\ref{fig:CF_twopole}, one can clearly observe the distinct behavior of the two poles in different final states. 
In the left panel, the $\rho\bar{K}$ contribution is primarily governed by the higher pole, so that the red line exhibits a typical resonant behavior. 
In contrast, the $\bar{K}^*\pi$ contribution is mainly dominated by the lower pole, where the blue line gradually increases from zero and approaches unity, showing a clear signature of a deeply bound state relative to the $\rho^-\bar{K}^0$ threshold~\cite{Liu:2024nac}. 
In the right panel, the situation is essentially reversed: the blue line almost entirely dominates the low-momentum region of the total CFs shown in Fig.~\ref{fig:CFQm1}, simply because the lower pole plays a crucial role in the $\bar{K}^*\pi$ component.

\section{Conclusion and outlook}
\label{sec:summary}
In this work, we have performed a comprehensive study of the axial-vector resonance $K_1(1270)$ within the two-pole scenario predicted by unitarized chiral perturbation theory~(UChPT).  
Using a coupled-channel framework including the dominant $\rho K$ and $K^*\pi$ interactions, we reproduced the well-known two-pole structure originally identified in Refs.~\cite{Geng:2006yb,Xie:2023cej} and investigated its dependence on the light-quark masses.  
The fitted pion-mass evolution of the $\rho$ and $K^*$ vector mesons, constrained by both lattice-QCD and experimental data, enabled us to trace the trajectories of the two poles in the complex-energy plane with increasing pion mass.  
We further examined the influence of vector-meson decay widths by comparing calculations with and without width effects, demonstrating that including finite widths smooths the pole evolution but does not alter the qualitative two-pole pattern.

The lower pole, dominantly coupled to the $K^*\pi$ channel, evolves from an above-threshold resonance to a bound or virtual state as the pion mass increases, 
while the higher pole, mainly associated with $\rho K$, moves downward in energy, showing a strengthened attraction consistent with chiral expectations.  
These results establish a consistent and physically intuitive picture of the chiral evolution of the $K_1(1270)$ two-pole structure.

In addition, we have carried out a detailed analysis of the femtoscopic CFs for all relevant vector-pseudoscalar~(VP) channels in both charge sectors.  
The calculated CFs clearly reflect the two-pole dynamics and exhibit distinct resonance and bound-state features in different channels.  
We have also shown that the vector meson finite widths play a smoothing role in the CF line shapes, and that the Coulomb interaction produces sizable low-momentum enhancements in the charged channels $\rho^+K^-$ and $K^{*-}\pi^+$.  
Moreover, the negligible influence of the higher channels such as $\omega \bar{K}$, $\bar{K}^*\eta$, and $\phi\bar{K}$ further confirms that the simplified two-channel description ($\rho K$, $K^*\pi$) captures the essential dynamics of the $K_1(1270)$ system.

Future directions include the extension of the present framework to incorporate additional coupled channels with improved lattice QCD constraints, 
as well as the direct comparison of the predicted CFs with experimental measurements from ALICE and future facilities.  
Such developments will allow a more precise determination of the $K_1(1270)$ pole parameters and a deeper understanding of its molecular components, 
offering valuable insight into the nonperturbative dynamics of chiral symmetry breaking and resonance formation in QCD.

%%%%%%%%%%%%%%%%%%%%%%%%%%%%
%%%%%%%%%%%%%%%%%%%%%%%%%%%%

\textit{Acknowledgment}.--- This work is partly supported by the National Key R\&D Program of China under Grant No. 2023YFA1606703 and the National Natural Science Foundation of China under Grant No.12435007  and No.12105006. Z.W.L acknowledges support from the National Natural Science Foundation of China under Grant No.12405133, No.12347180, China Postdoctoral Science Foundation under Grant No.2023M740189, and the Postdoctoral Fellowship Program of CPSF under Grant No.GZC20233381. H.L. acknowledges the JSPS Grant-in-Aid for Scientific Research (S) under Grant No. JP20H05648. R. M. acknowledges support from the ESGENT program (ESGENT/018/2024) and the PROMETEU program (CIPROM/2023/59), of the Generalitat Valenciana, and also from the Spanish Ministerio de
Economia y Competitividad (MINECO) and European
Union (NextGenerationEU/PRTR) by the grants with
Ref. CNS2022-136146, and Ref. PID2023-147458NB-C21.

\bibliography{bib.bib}

@article{Molina:2020qpw,
    author = "Molina, R. and Ruiz de Elvira, J.",
    title = "{Light- and strange-quark mass dependence of the $\rho$(770) meson revisited}",
    eprint = "2005.13584",
    archivePrefix = "arXiv",
    primaryClass = "hep-lat",
    doi = "10.1007/JHEP11(2020)017",
    journal = "JHEP",
    volume = "11",
    pages = "017",
    year = "2020"
}

@article{Guo:2016zos,
    author = {Guo, Dehua and Alexandru, Andrei and Molina, Raquel and D{\"o}ring, Michael},
    title = "{Rho resonance parameters from lattice QCD}",
    eprint = "1605.03993",
    archivePrefix = "arXiv",
    primaryClass = "hep-lat",
    reportNumber = "JLAB-THY-16-2313",
    doi = "10.1103/PhysRevD.94.034501",
    journal = "Phys. Rev. D",
    volume = "94",
    number = "3",
    pages = "034501",
    year = "2016"
}

@article{Hanhart:2008mx,
    author = "Hanhart, C. and Pelaez, J. R. and Rios, G.",
    title = "{Quark mass dependence of the rho and sigma from dispersion relations and Chiral Perturbation Theory}",
    eprint = "0801.2871",
    archivePrefix = "arXiv",
    primaryClass = "hep-ph",
    reportNumber = "FZJ-IKP-TH-2008-01",
    doi = "10.1103/PhysRevLett.100.152001",
    journal = "Phys. Rev. Lett.",
    volume = "100",
    pages = "152001",
    year = "2008"
}

@article{GomezNicola:2001as,
    author = "Gomez Nicola, A. and Pelaez, J. R.",
    title = "{Meson meson scattering within one loop chiral perturbation theory and its unitarization}",
    eprint = "hep-ph/0109056",
    archivePrefix = "arXiv",
    doi = "10.1103/PhysRevD.65.054009",
    journal = "Phys. Rev. D",
    volume = "65",
    pages = "054009",
    year = "2002"
}

@article{Bruns:2004tj,
    author = "Bruns, Peter C. and Meissner, Ulf-G.",
    title = "{Infrared regularization for spin-1 fields}",
    eprint = "hep-ph/0411223",
    archivePrefix = "arXiv",
    reportNumber = "HISKP-TH-04-23",
    doi = "10.1140/epjc/s2005-02118-0",
    journal = "Eur. Phys. J. C",
    volume = "40",
    pages = "97--119",
    year = "2005"
}

@article{Liu:2025rci,
    author = "Liu, Zhi-Wei and Lu, Jun-Xu and Geng, Li-Sheng",
    title = "{The femtoscopic technique{\textemdash}an invaluable tool in studies of exotic hadrons}",
    doi = "10.22323/1.465.0044",
    journal = "PoS",
    volume = "QNP2024",
    pages = "044",
    year = "2025"
}

@article{Shen:2025qpj,
    author = "Shen, Yi-bo and Liu, Zhi-Wei and Lu, Jun-Xu and Liu, Ming-Zhu and Geng, Li-Sheng",
    title = "{Probing the structure of exotic hadrons through correlation functions}",
    eprint = "2506.23476",
    archivePrefix = "arXiv",
    primaryClass = "hep-ph",
    month = "6",
    year = "2025"
}

@article{Zhuang:2024udv,
    author = "Zhuang, Zejian and Molina, Raquel and Lu, Jun-Xu and Geng, Li-Sheng",
    title = "{Pole trajectories of the {\ensuremath{\Lambda}}(1405) help establish its dynamical nature}",
    eprint = "2405.07686",
    archivePrefix = "arXiv",
    primaryClass = "hep-ph",
    doi = "10.1016/j.scib.2025.04.029",
    journal = "Sci. Bull.",
    volume = "70",
    pages = "1953--1961",
    year = "2025"
}

@article{Feng:2010es,
    author = "Feng, Xu and Jansen, Karl and Renner, Dru B.",
    title = "{Resonance Parameters of the rho-Meson from Lattice QCD}",
    eprint = "1011.5288",
    archivePrefix = "arXiv",
    primaryClass = "hep-lat",
    reportNumber = "DESY-10-176, SFB-CPP-10-117, MS-TP-10-14, KEK-CP-242, JLAB-THY-10-1289",
    doi = "10.1103/PhysRevD.83.094505",
    journal = "Phys. Rev. D",
    volume = "83",
    pages = "094505",
    year = "2011"
}

@article{Yu:2023xxf,
    author = "Yu, Kang and Li, Yan and Wu, Jia-Jun and Leinweber, Derek B. and Thomas, Anthony W.",
    title = "{Study of the pion-mass dependence of {\ensuremath{\rho}}-meson properties in lattice QCD}",
    eprint = "2311.03903",
    archivePrefix = "arXiv",
    primaryClass = "hep-lat",
    reportNumber = "ADP-23-27/T1236",
    doi = "10.1103/PhysRevD.109.034505",
    journal = "Phys. Rev. D",
    volume = "109",
    number = "3",
    pages = "034505",
    year = "2024"
}

@article{Koonin:1977fh,
    author = "Koonin, S. E.",
    title = "{Proton Pictures of High-Energy Nuclear Collisions}",
    doi = "10.1016/0370-2693(77)90340-9",
    journal = "Phys. Lett. B",
    volume = "70",
    pages = "43--47",
    year = "1977"
}

@article{Pratt:1990zq,
    author = "Pratt, S. and Csorgo, T. and Zimanyi, J.",
    title = "{Detailed predictions for two pion correlations in ultrarelativistic heavy ion collisions}",
    doi = "10.1103/PhysRevC.42.2646",
    journal = "Phys. Rev. C",
    volume = "42",
    pages = "2646--2652",
    year = "1990"
}

@article{Wilson:2019wfr,
    author = "Wilson, David J. and Briceno, Raul A. and Dudek, Jozef J. and Edwards, Robert G. and Thomas, Christopher E.",
    title = "{The quark-mass dependence of elastic $\pi K$ scattering from QCD}",
    eprint = "1904.03188",
    archivePrefix = "arXiv",
    primaryClass = "hep-lat",
    reportNumber = "DAMTP-2019-13, JLAB-THY-19-2911",
    doi = "10.1103/PhysRevLett.123.042002",
    journal = "Phys. Rev. Lett.",
    volume = "123",
    number = "4",
    pages = "042002",
    year = "2019"
}

@article{Wilson:2014cna,
    author = "Wilson, David J. and Dudek, Jozef J. and Edwards, Robert G. and Thomas, Christopher E.",
    title = "{Resonances in coupled $\pi K, \eta K$ scattering from lattice QCD}",
    eprint = "1411.2004",
    archivePrefix = "arXiv",
    primaryClass = "hep-ph",
    reportNumber = "JLAB-THY-14-1982, DAMTP-2014-81, JLAB-THY-14-1892",
    doi = "10.1103/PhysRevD.91.054008",
    journal = "Phys. Rev. D",
    volume = "91",
    number = "5",
    pages = "054008",
    year = "2015"
}

@article{Wang:2025hew,
    author = "Wang, Zhengli and Leinweber, Derek B. and Liu, Chuan and Liu, Liuming and Sun, Peng and Thomas, Anthony W. and Wu, Jia-jun and Xing, Hanyang and Yu, Kang",
    collaboration = "CLQCD",
    title = "{Spectral parameters of the {\ensuremath{\rho}} resonance from lattice QCD}",
    eprint = "2502.03700",
    archivePrefix = "arXiv",
    primaryClass = "hep-lat",
    reportNumber = "ADP-25-4/1266",
    doi = "10.1007/JHEP08(2025)064",
    journal = "JHEP",
    volume = "08",
    pages = "064",
    year = "2025"
}

@article{Rodas:2023gma,
    author = "Rodas, Arkaitz and Dudek, Jozef J. and Edwards, Robert G.",
    collaboration = "Hadron Spectrum",
    title = "{Quark mass dependence of {\ensuremath{\pi}}{\ensuremath{\pi}} scattering in isospin 0, 1, and 2 from lattice QCD}",
    eprint = "2303.10701",
    archivePrefix = "arXiv",
    primaryClass = "hep-lat",
    reportNumber = "JLAB-THY-23-3778",
    doi = "10.1103/PhysRevD.108.034513",
    journal = "Phys. Rev. D",
    volume = "108",
    number = "3",
    pages = "034513",
    year = "2023"
}

@article{Wilson:2015dqa,
    author = "Wilson, David J. and Briceno, Raul A. and Dudek, Jozef J. and Edwards, Robert G. and Thomas, Christopher E.",
    title = "{Coupled $\pi\pi, K\bar{K}$ scattering in $P$-wave and the $\rho$ resonance from lattice QCD}",
    eprint = "1507.02599",
    archivePrefix = "arXiv",
    primaryClass = "hep-ph",
    reportNumber = "DAMTP-2015-34, JLAB-THY-15-2101",
    doi = "10.1103/PhysRevD.92.094502",
    journal = "Phys. Rev. D",
    volume = "92",
    number = "9",
    pages = "094502",
    year = "2015"
}

@article{Dudek:2012xn,
    author = "Dudek, Jozef J. and Edwards, Robert G. and Thomas, Christopher E.",
    collaboration = "Hadron Spectrum",
    title = "{Energy dependence of the $\rho$ resonance in $\pi\pi$ elastic scattering from lattice QCD}",
    eprint = "1212.0830",
    archivePrefix = "arXiv",
    primaryClass = "hep-ph",
    reportNumber = "JLAB-THY-12-1666, TCDMATH-12-10",
    doi = "10.1103/PhysRevD.87.034505",
    journal = "Phys. Rev. D",
    volume = "87",
    number = "3",
    pages = "034505",
    year = "2013",
    note = "[Erratum: Phys.Rev.D 90, 099902 (2014)]"
}

@article{Hyodo:2013iga,
    author = "Hyodo, Tetsuo",
    title = "{Structure of Near-Threshold s-Wave Resonances}",
    eprint = "1305.1999",
    archivePrefix = "arXiv",
    primaryClass = "hep-ph",
    doi = "10.1103/PhysRevLett.111.132002",
    journal = "Phys. Rev. Lett.",
    volume = "111",
    pages = "132002",
    year = "2013"
}

@article{Ramos:2025ibe,
    author = "Ramos, {\`A}ngels and Torres-Rincon, Juan M. and de Fagoaga, Alejandro and Cabr{\'e}, Esteve",
    title = "{Kaon-deuteron femtoscopy from unitarized chiral interactions}",
    eprint = "2507.22593",
    archivePrefix = "arXiv",
    primaryClass = "hep-ph",
    month = "7",
    year = "2025"
}

@article{Encarnacion:2024jge,
    author = "Encarnaci{\'o}n, P. and Feijoo, A. and Sarti, V. Mantovani and Ramos, A.",
    title = "{Femtoscopic study of the S=-1 meson-baryon interaction: K-p, {\ensuremath{\pi}}-{\ensuremath{\Lambda}}, and K+{\ensuremath{\Xi}}- correlations}",
    eprint = "2412.20880",
    archivePrefix = "arXiv",
    primaryClass = "hep-ph",
    doi = "10.1103/3ycr-vzmd",
    journal = "Phys. Rev. D",
    volume = "111",
    number = "11",
    pages = "114013",
    year = "2025"
}

@article{Feijoo:2024bvn,
    author = "Feijoo, A. and Korwieser, M. and Fabbietti, L.",
    title = "{Relevance of the coupled channels in the {\ensuremath{\phi}}p and {\ensuremath{\rho}}0p correlation functions}",
    eprint = "2407.01128",
    archivePrefix = "arXiv",
    primaryClass = "hep-ph",
    doi = "10.1103/PhysRevD.111.014009",
    journal = "Phys. Rev. D",
    volume = "111",
    number = "1",
    pages = "014009",
    year = "2025"
}

@article{Sarti:2023wlg,
    author = "Sarti, V. Mantovani and Feijoo, A. and Vida{\~n}a, I. and Ramos, A. and Giacosa, F. and Hyodo, T. and Kamiya, Y.",
    title = "{Constraining the low-energy S=-2 meson-baryon interaction with two-particle correlations}",
    eprint = "2309.08756",
    archivePrefix = "arXiv",
    primaryClass = "hep-ph",
    doi = "10.1103/PhysRevD.110.L011505",
    journal = "Phys. Rev. D",
    volume = "110",
    number = "1",
    pages = "L011505",
    year = "2024"
}

@article{Albaladejo:2024lam,
    author = "Albaladejo, M. and Feijoo, A. and Nieves, J. and Oset, E. and Vida{\~n}a, I.",
    title = "{Femtoscopy correlation functions and mass distributions from production experiments}",
    eprint = "2410.08880",
    archivePrefix = "arXiv",
    primaryClass = "hep-ph",
    doi = "10.1103/PhysRevD.110.114052",
    journal = "Phys. Rev. D",
    volume = "110",
    number = "11",
    pages = "114052",
    year = "2024"
}

@article{Torres-Rincon:2023qll,
    author = "Torres-Rincon, Juan M. and Ramos, {\`A}ngels and Tolos, Laura",
    title = "{Femtoscopy of D mesons and light mesons upon unitarized effective field theories}",
    eprint = "2307.02102",
    archivePrefix = "arXiv",
    primaryClass = "hep-ph",
    doi = "10.1103/PhysRevD.108.096008",
    journal = "Phys. Rev. D",
    volume = "108",
    number = "9",
    pages = "096008",
    year = "2023"
}

@article{Ikeno:2023ojl,
    author = "Ikeno, Natsumi and Toledo, Genaro and Oset, Eulogio",
    title = "{Model independent analysis of femtoscopic correlation functions: An application to the Ds0{\textasteriskcentered}(2317)}",
    eprint = "2305.16431",
    archivePrefix = "arXiv",
    primaryClass = "hep-ph",
    doi = "10.1016/j.physletb.2023.138281",
    journal = "Phys. Lett. B",
    volume = "847",
    pages = "138281",
    year = "2023"
}

@article{Vidana:2023olz,
    author = "Vidana, I. and Feijoo, A. and Albaladejo, M. and Nieves, J. and Oset, E.",
    title = "{Femtoscopic correlation function for the Tcc(3875)+ state}",
    eprint = "2303.06079",
    archivePrefix = "arXiv",
    primaryClass = "hep-ph",
    doi = "10.1016/j.physletb.2023.138201",
    journal = "Phys. Lett. B",
    volume = "846",
    pages = "138201",
    year = "2023"
}

@article{Liu:2025oar,
    author = "Liu, Zhi-Wei and Ge, Duo-Lun and Lu, Jun-Xu and Liu, Ming-Zhu and Geng, Li-Sheng",
    title = "{Charmonium-nucleon femtoscopic correlation function}",
    eprint = "2504.04853",
    archivePrefix = "arXiv",
    primaryClass = "hep-ph",
    doi = "10.1103/3bdh-blwh",
    journal = "Phys. Rev. D",
    volume = "112",
    number = "5",
    pages = "054019",
    year = "2025"
}

@article{Ge:2025put,
    author = "Ge, Duo-Lun and Liu, Zhi-Wei and Lu, Jun-Xu and Geng, Li-Sheng",
    title = "{Deuteron-deuteron interaction and correlation function}",
    eprint = "2502.18872",
    archivePrefix = "arXiv",
    primaryClass = "nucl-th",
    doi = "10.1103/ttrc-qhv5",
    journal = "Phys. Rev. C",
    volume = "112",
    number = "3",
    pages = "034003",
    year = "2025"
}

@article{Molina:2023oeu,
    author = "Molina, R. and Liu, Zhi-Wei and Geng, Li-Sheng and Oset, E.",
    title = "{Correlation function for the $a_0(980)$}",
    eprint = "2312.11993",
    archivePrefix = "arXiv",
    primaryClass = "hep-ph",
    doi = "10.1140/epjc/s10052-024-12694-w",
    journal = "Eur. Phys. J. C",
    volume = "84",
    number = "3",
    pages = "328",
    year = "2024"
}

@article{Liu:2022nec,
    author = "Liu, Zhi-Wei and Li, Kai-Wen and Geng, Li-Sheng",
    title = "{Strangeness S = {\ensuremath{-}}2 baryon-baryon interactions and femtoscopic correlation functions in covariant chiral effective field theory*}",
    eprint = "2201.04997",
    archivePrefix = "arXiv",
    primaryClass = "hep-ph",
    doi = "10.1088/1674-1137/ac988a",
    journal = "Chin. Phys. C",
    volume = "47",
    number = "2",
    pages = "024108",
    year = "2023"
}

@article{Liu:2023wfo,
    author = "Liu, Zhi-Wei and Lu, Jun-Xu and Liu, Ming-Zhu and Geng, Li-Sheng",
    title = "{Distinguishing the spins of Pc(4440) and Pc(4457) with femtoscopic correlation functions}",
    eprint = "2305.19048",
    archivePrefix = "arXiv",
    primaryClass = "hep-ph",
    doi = "10.1103/PhysRevD.108.L031503",
    journal = "Phys. Rev. D",
    volume = "108",
    number = "3",
    pages = "L031503",
    year = "2023"
}

@article{Liu:2024uxn,
    author = "Liu, Ming-Zhu and Pan, Ya-Wen and Liu, Zhi-Wei and Wu, Tian-Wei and Lu, Jun-Xu and Geng, Li-Sheng",
    title = "{Three ways to decipher the nature of exotic hadrons: Multiplets, three-body hadronic molecules, and correlation functions}",
    eprint = "2404.06399",
    archivePrefix = "arXiv",
    primaryClass = "hep-ph",
    doi = "10.1016/j.physrep.2024.12.001",
    journal = "Phys. Rept.",
    volume = "1108",
    pages = "1--108",
    year = "2025"
}

@article{Liu:2024nac,
    author = "Liu, Zhi-Wei and Lu, Jun-Xu and Liu, Ming-Zhu and Geng, Li-Sheng",
    title = "{Femtoscopy can tell whether $Z_c(3900)$ and $Z_{cs}(3985)$ are resonances or virtual states}",
    eprint = "2404.18607",
    archivePrefix = "arXiv",
    primaryClass = "hep-ph",
    month = "4",
    year = "2024"
}

@article{Liu:2023uly,
    author = "Liu, Zhi-Wei and Lu, Jun-Xu and Geng, Li-Sheng",
    title = "{Study of the DK interaction with femtoscopic correlation functions}",
    eprint = "2302.01046",
    archivePrefix = "arXiv",
    primaryClass = "hep-ph",
    doi = "10.1103/PhysRevD.107.074019",
    journal = "Phys. Rev. D",
    volume = "107",
    number = "7",
    pages = "074019",
    year = "2023"
}

@article{Kamiya:2019uiw,
    author = "Kamiya, Yuki and Hyodo, Tetsuo and Morita, Kenji and Ohnishi, Akira and Weise, Wolfram",
    title = "{$K^-p$ Correlation Function from High-Energy Nuclear Collisions and Chiral SU(3) Dynamics}",
    eprint = "1911.01041",
    archivePrefix = "arXiv",
    primaryClass = "nucl-th",
    doi = "10.1103/PhysRevLett.124.132501",
    journal = "Phys. Rev. Lett.",
    volume = "124",
    number = "13",
    pages = "132501",
    year = "2020"
}

@article{BaryonScatteringBaSc:2023zvt,
    author = "Bulava, John and others",
    collaboration = "Baryon Scattering (BaSc)",
    title = "{Two-Pole Nature of the \ensuremath{\Lambda}(1405) resonance from Lattice QCD}",
    eprint = "2307.10413",
    archivePrefix = "arXiv",
    primaryClass = "hep-lat",
    reportNumber = "MIT-CTP/5579",
    doi = "10.1103/PhysRevLett.132.051901",
    journal = "Phys. Rev. Lett.",
    volume = "132",
    number = "5",
    pages = "051901",
    year = "2024"
}

@article{BaryonScatteringBaSc:2023ori,
    author = "Bulava, John and others",
    collaboration = "Baryon Scattering (BaSc)",
    title = "{Lattice QCD study of \ensuremath{\pi}\ensuremath{\Sigma}-K\textasciimacron{}N scattering and the \ensuremath{\Lambda}(1405) resonance}",
    eprint = "2307.13471",
    archivePrefix = "arXiv",
    primaryClass = "hep-lat",
    reportNumber = "MIT-CTP/5580",
    doi = "10.1103/PhysRevD.109.014511",
    journal = "Phys. Rev. D",
    volume = "109",
    number = "1",
    pages = "014511",
    year = "2024"
}

@article{Briceno:2017max,
    author = "Briceno, Raul A. and Dudek, Jozef J. and Young, Ross D.",
    title = "{Scattering processes and resonances from lattice QCD}",
    eprint = "1706.06223",
    archivePrefix = "arXiv",
    primaryClass = "hep-lat",
    reportNumber = "JLAB-THY-17-2495, ADP-17-28-T1034",
    doi = "10.1103/RevModPhys.90.025001",
    journal = "Rev. Mod. Phys.",
    volume = "90",
    number = "2",
    pages = "025001",
    year = "2018"
}

@inproceedings{Xie:2025nnq,
    author = "Xie, Jia-Ming and Lu, Jun-Xu and Geng, Li-Sheng and Zou, Bing-Song",
    title = "{Two-pole structures in QCD -- a universal phenomenon governed by chiral dynamics}",
    booktitle = "{11th International Workshop on Chiral Dynamics}",
    eprint = "2504.03392",
    archivePrefix = "arXiv",
    primaryClass = "hep-ph",
    month = "4",
    year = "2025"
}

@article{Xie:2023jve,
    author = "Xie, Jia-Ming and Lu, Jun-Xu and Geng, Li-Sheng and Zou, Bing-Song",
    title = "{Dynamical origin of universal two-pole structures and their light quark mass evolution}",
    eprint = "2312.17287",
    archivePrefix = "arXiv",
    primaryClass = "hep-ph",
    doi = "10.1051/epjconf/202430301011",
    journal = "EPJ Web Conf.",
    volume = "303",
    pages = "01011",
    year = "2024"
}

@article{Xie:2023cej,
    author = "Xie, Jia-Ming and Lu, Jun-Xu and Geng, Li-Sheng and Zou, Bing-Song",
    title = "{Two-pole structures as a universal phenomenon dictated by coupled-channel chiral dynamics}",
    eprint = "2307.11631",
    archivePrefix = "arXiv",
    primaryClass = "hep-ph",
    doi = "10.1103/PhysRevD.108.L111502",
    journal = "Phys. Rev. D",
    volume = "108",
    number = "11",
    pages = "L111502",
    year = "2023"
}

@article{ParticleDataGroup:2024cfk,
    author = "Navas, S. and others",
    collaboration = "Particle Data Group",
    title = "{Review of particle physics}",
    doi = "10.1103/PhysRevD.110.030001",
    journal = "Phys. Rev. D",
    volume = "110",
    number = "3",
    pages = "030001",
    year = "2024"
}

@article{Weinberg:1990rz,
    author = "Weinberg, Steven",
    title = "{Nuclear forces from chiral Lagrangians}",
    reportNumber = "UTTG-31-90",
    doi = "10.1016/0370-2693(90)90938-3",
    journal = "Phys. Lett. B",
    volume = "251",
    pages = "288--292",
    year = "1990"
}

@article{Kaiser:1996js,
    author = "Kaiser, Norbert and Waas, T. and Weise, W.",
    title = "{SU(3) chiral dynamics with coupled channels: Eta and kaon photoproduction}",
    eprint = "hep-ph/9607459",
    archivePrefix = "arXiv",
    doi = "10.1016/S0375-9474(96)00321-1",
    journal = "Nucl. Phys. A",
    volume = "612",
    pages = "297--320",
    year = "1997"
}

@article{Roca:2021bxk,
    author = "Roca, L. and Liang, W. H. and Oset, E.",
    title = "{Inconsistency of the data on the K1(1270){\textrightarrow}{\ensuremath{\pi}}K0{\textasteriskcentered}(1430) decay width}",
    eprint = "2105.11768",
    archivePrefix = "arXiv",
    primaryClass = "hep-ph",
    doi = "10.1016/j.physletb.2021.136827",
    journal = "Phys. Lett. B",
    volume = "824",
    pages = "136827",
    year = "2022"
}

@article{Dias:2021upl,
    author = "Dias, J. M. and Toledo, G. and Roca, L. and Oset, E.",
    title = "{Unveiling the K1(1270) double-pole structure in the B{\textasciimacron}{\textrightarrow}J/{\ensuremath{\psi}}{\ensuremath{\rho}}K{\textasciimacron} and B{\textasciimacron}{\textrightarrow}J/{\ensuremath{\psi}}K{\textasciimacron}*{\ensuremath{\pi}} decays}",
    eprint = "2102.08402",
    archivePrefix = "arXiv",
    primaryClass = "hep-ph",
    doi = "10.1103/PhysRevD.103.116019",
    journal = "Phys. Rev. D",
    volume = "103",
    number = "11",
    pages = "116019",
    year = "2021"
}

@article{Wang:2020pyy,
    author = "Wang, Guan-Ying and Roca, Luis and Wang, En and Liang, Wei-Hong and Oset, Eulogio",
    title = "{Signatures of the two $K_1(1270)$ poles in $D^+\to \nu e^+ V P$ decay}",
    eprint = "2002.07610",
    archivePrefix = "arXiv",
    primaryClass = "hep-ph",
    doi = "10.1140/epjc/s10052-020-7939-1",
    journal = "Eur. Phys. J. C",
    volume = "80",
    number = "5",
    pages = "388",
    year = "2020"
}

@article{Wang:2019mph,
    author = "Wang, G. Y. and Roca, L. and Oset, E.",
    title = "{Discerning the two $K_1(1270)$ poles in $D^0\to \pi^+ V P$ decay}",
    eprint = "1907.09188",
    archivePrefix = "arXiv",
    primaryClass = "hep-ph",
    doi = "10.1103/PhysRevD.100.074018",
    journal = "Phys. Rev. D",
    volume = "100",
    number = "7",
    pages = "074018",
    year = "2019"
}

@article{Lutz:2003fm,
    author = "Lutz, M. F. M. and Kolomeitsev, E. E.",
    title = "{On meson resonances and chiral symmetry}",
    eprint = "nucl-th/0307039",
    archivePrefix = "arXiv",
    reportNumber = "GSI-PREPRINT-2003-19",
    doi = "10.1016/j.nuclphysa.2003.11.009",
    journal = "Nucl. Phys. A",
    volume = "730",
    pages = "392--416",
    year = "2004"
}

@article{Brambilla:2010cs,
    author = "Brambilla, N. and others",
    title = "{Heavy Quarkonium: Progress, Puzzles, and Opportunities}",
    eprint = "1010.5827",
    archivePrefix = "arXiv",
    primaryClass = "hep-ph",
    reportNumber = "SLAC-R-996, TUM-EFT-11-10, CLNS-10-2066, ANL-HEP-PR-10-44, ALBERTA-THY-11-10, CP3-10-37, FZJ-IKP-TH-2010-24, INT-PUB-10-059, JLAB-THY-11-1308, FERMILAB-PUB-10-737-T",
    doi = "10.1140/epjc/s10052-010-1534-9",
    journal = "Eur. Phys. J. C",
    volume = "71",
    pages = "1534",
    year = "2011"
}

@article{Lebed:2016hpi,
    author = "Lebed, Richard F. and Mitchell, Ryan E. and Swanson, Eric S.",
    title = "{Heavy-Quark QCD Exotica}",
    eprint = "1610.04528",
    archivePrefix = "arXiv",
    primaryClass = "hep-ph",
    doi = "10.1016/j.ppnp.2016.11.003",
    journal = "Prog. Part. Nucl. Phys.",
    volume = "93",
    pages = "143--194",
    year = "2017"
}

@article{Hosaka:2016ypm,
    author = "Hosaka, Atsushi and Hyodo, Tetsuo and Sudoh, Kazutaka and Yamaguchi, Yasuhiro and Yasui, Shigehiro",
    title = "{Heavy Hadrons in Nuclear Matter}",
    eprint = "1606.08685",
    archivePrefix = "arXiv",
    primaryClass = "hep-ph",
    reportNumber = "J-PARC-TH-0055, YITP-16-82",
    doi = "10.1016/j.ppnp.2017.04.003",
    journal = "Prog. Part. Nucl. Phys.",
    volume = "96",
    pages = "88--153",
    year = "2017"
}

@article{Richard:2016eis,
    author = "Richard, Jean-Marc",
    title = "{Exotic hadrons: review and perspectives}",
    eprint = "1606.08593",
    archivePrefix = "arXiv",
    primaryClass = "hep-ph",
    doi = "10.1007/s00601-016-1159-0",
    journal = "Few Body Syst.",
    volume = "57",
    number = "12",
    pages = "1185--1212",
    year = "2016"
}

@article{Chen:2022asf,
    author = "Chen, Hua-Xing and Chen, Wei and Liu, Xiang and Liu, Yan-Rui and Zhu, Shi-Lin",
    title = "{An updated review of the new hadron states}",
    eprint = "2204.02649",
    archivePrefix = "arXiv",
    primaryClass = "hep-ph",
    doi = "10.1088/1361-6633/aca3b6",
    journal = "Rept. Prog. Phys.",
    volume = "86",
    number = "2",
    pages = "026201",
    year = "2023"
}

@article{Liu:2019zoy,
    author = "Liu, Yan-Rui and Chen, Hua-Xing and Chen, Wei and Liu, Xiang and Zhu, Shi-Lin",
    title = "{Pentaquark and Tetraquark states}",
    eprint = "1903.11976",
    archivePrefix = "arXiv",
    primaryClass = "hep-ph",
    doi = "10.1016/j.ppnp.2019.04.003",
    journal = "Prog. Part. Nucl. Phys.",
    volume = "107",
    pages = "237--320",
    year = "2019"
}

@article{Karliner:2017qhf,
    author = "Karliner, Marek and Rosner, Jonathan L. and Skwarnicki, Tomasz",
    title = "{Multiquark States}",
    eprint = "1711.10626",
    archivePrefix = "arXiv",
    primaryClass = "hep-ph",
    doi = "10.1146/annurev-nucl-101917-020902",
    journal = "Ann. Rev. Nucl. Part. Sci.",
    volume = "68",
    pages = "17--44",
    year = "2018"
}

@article{Olsen:2017bmm,
    author = "Olsen, Stephen Lars and Skwarnicki, Tomasz and Zieminska, Daria",
    title = "{Nonstandard heavy mesons and baryons: Experimental evidence}",
    eprint = "1708.04012",
    archivePrefix = "arXiv",
    primaryClass = "hep-ph",
    doi = "10.1103/RevModPhys.90.015003",
    journal = "Rev. Mod. Phys.",
    volume = "90",
    number = "1",
    pages = "015003",
    year = "2018"
}

@article{Esposito:2016noz,
    author = "Esposito, A. and Pilloni, A. and Polosa, A. D.",
    title = "{Multiquark Resonances}",
    eprint = "1611.07920",
    archivePrefix = "arXiv",
    primaryClass = "hep-ph",
    reportNumber = "JLAB-THY-16-2301",
    doi = "10.1016/j.physrep.2016.11.002",
    journal = "Phys. Rept.",
    volume = "668",
    pages = "1--97",
    year = "2017"
}

@article{Chen:2016qju,
    author = "Chen, Hua-Xing and Chen, Wei and Liu, Xiang and Zhu, Shi-Lin",
    title = "{The hidden-charm pentaquark and tetraquark states}",
    eprint = "1601.02092",
    archivePrefix = "arXiv",
    primaryClass = "hep-ph",
    doi = "10.1016/j.physrep.2016.05.004",
    journal = "Phys. Rept.",
    volume = "639",
    pages = "1--121",
    year = "2016"
}

@article{Birse:1996hd,
    author = "Birse, Michael C.",
    title = "{Effective chiral Lagrangians for spin 1 mesons}",
    eprint = "hep-ph/9603251",
    archivePrefix = "arXiv",
    reportNumber = "MC-TH-96-11",
    doi = "10.1007/s002180050105",
    journal = "Z. Phys. A",
    volume = "355",
    pages = "231--246",
    year = "1996"
}

@article{Hyodo:2011ur,
    author = "Hyodo, Tetsuo and Jido, Daisuke",
    title = "{The nature of the Lambda(1405) resonance in chiral dynamics}",
    eprint = "1104.4474",
    archivePrefix = "arXiv",
    primaryClass = "nucl-th",
    doi = "10.1016/j.ppnp.2011.07.002",
    journal = "Prog. Part. Nucl. Phys.",
    volume = "67",
    pages = "55--98",
    year = "2012"
}

@article{Oller:2019opk,
    author = "Oller, J. A.",
    title = "{Coupled-channel approach in hadron\textendash{}hadron scattering}",
    eprint = "1909.00370",
    archivePrefix = "arXiv",
    primaryClass = "hep-ph",
    doi = "10.1016/j.ppnp.2019.103728",
    journal = "Prog. Part. Nucl. Phys.",
    volume = "110",
    pages = "103728",
    year = "2020"
}

@article{Roca:2005nm,
    author = "Roca, L. and Oset, E. and Singh, J.",
    title = "{Low lying axial-vector mesons as dynamically generated resonances}",
    eprint = "hep-ph/0503273",
    archivePrefix = "arXiv",
    doi = "10.1103/PhysRevD.72.014002",
    journal = "Phys. Rev. D",
    volume = "72",
    pages = "014002",
    year = "2005"
}

@article{Mai:2022eur,
    author = "Mai, Maxim and Mei\ss{}ner, Ulf-G. and Urbach, Carsten",
    title = "{Towards a theory of hadron resonances}",
    eprint = "2206.01477",
    archivePrefix = "arXiv",
    primaryClass = "hep-ph",
    doi = "10.1016/j.physrep.2022.11.005",
    journal = "Phys. Rept.",
    volume = "1001",
    pages = "1--66",
    year = "2023"
}

@article{Jido:2003cb,
    author = "Jido, D. and Oller, J. A. and Oset, E. and Ramos, A. and Meissner, U. G.",
    title = "{Chiral dynamics of the two Lambda(1405) states}",
    eprint = "nucl-th/0303062",
    archivePrefix = "arXiv",
    doi = "10.1016/S0375-9474(03)01598-7",
    journal = "Nucl. Phys. A",
    volume = "725",
    pages = "181--200",
    year = "2003"
}

@article{Godfrey:1985xj,
    author = "Godfrey, S. and Isgur, Nathan",
    title = "{Mesons in a Relativized Quark Model with Chromodynamics}",
    doi = "10.1103/PhysRevD.32.189",
    journal = "Phys. Rev. D",
    volume = "32",
    pages = "189--231",
    year = "1985"
}

@article{Oller:1998hw,
    author = "Oller, J. A. and Oset, E. and Pelaez, J. R.",
    title = "{Meson meson interaction in a nonperturbative chiral approach}",
    eprint = "hep-ph/9804209",
    archivePrefix = "arXiv",
    reportNumber = "SLAC-PUB-7787",
    doi = "10.1103/PhysRevD.59.074001",
    journal = "Phys. Rev. D",
    volume = "59",
    pages = "074001",
    year = "1999",
    note = "[Erratum: Phys.Rev.D 60, 099906 (1999), Erratum: Phys.Rev.D 75, 099903 (2007)]"
}

@article{Oller:2000ma,
    author = "Oller, J. A. and Oset, E. and Ramos, A.",
    title = "{Chiral unitary approach to meson meson and meson - baryon interactions and nuclear applications}",
    eprint = "hep-ph/0002193",
    archivePrefix = "arXiv",
    reportNumber = "FZJ-IKP-TH-1999-37, FTUV-99-1215, IFIC-99-1215",
    doi = "10.1016/S0146-6410(00)00104-6",
    journal = "Prog. Part. Nucl. Phys.",
    volume = "45",
    pages = "157--242",
    year = "2000"
}

@article{Ren:2012aj,
    author = "Ren, X. -L. and Geng, L. S. and Martin Camalich, J. and Meng, J. and Toki, H.",
    title = "{Octet baryon masses in next-to-next-to-next-to-leading order covariant baryon chiral perturbation theory}",
    eprint = "1209.3641",
    archivePrefix = "arXiv",
    primaryClass = "nucl-th",
    doi = "10.1007/JHEP12(2012)073",
    journal = "JHEP",
    volume = "12",
    pages = "073",
    year = "2012"
}

@article{Geng:2006yb,
    author = "Geng, L. S. and Oset, E. and Roca, L. and Oller, J. A.",
    title = "{Clues for the existence of two K(1)(1270) resonances}",
    eprint = "hep-ph/0610217",
    archivePrefix = "arXiv",
    doi = "10.1103/PhysRevD.75.014017",
    journal = "Phys. Rev. D",
    volume = "75",
    pages = "014017",
    year = "2007"
}

@article{Oller:2000fj,
    author = "Oller, J. A. and Mei{\ss}ner, Ulf-G.",
    title = "{Chiral dynamics in the presence of bound states: Kaon nucleon interactions revisited}",
    eprint = "hep-ph/0011146",
    archivePrefix = "arXiv",
    reportNumber = "FZJ-IKP-TH-2000-26",
    doi = "10.1016/S0370-2693(01)00078-8",
    journal = "Phys. Lett. B",
    volume = "500",
    pages = "263--272",
    year = "2001"
}

@article{Oset:1997it,
    author = "Oset, E. and Ramos, A.",
    title = "{Nonperturbative chiral approach to s wave anti-K N interactions}",
    eprint = "nucl-th/9711022",
    archivePrefix = "arXiv",
    doi = "10.1016/S0375-9474(98)00170-5",
    journal = "Nucl. Phys. A",
    volume = "635",
    pages = "99--120",
    year = "1998"
}

@article{Kaiser:1995eg,
    author = "Kaiser, Norbert and Siegel, P. B. and Weise, W.",
    title = "{Chiral dynamics and the low-energy kaon - nucleon interaction}",
    eprint = "nucl-th/9505043",
    archivePrefix = "arXiv",
    reportNumber = "TUM-T39-95-5",
    doi = "10.1016/0375-9474(95)00362-5",
    journal = "Nucl. Phys. A",
    volume = "594",
    pages = "325--345",
    year = "1995"
}

@article{Guo:2017jvc,
    author = "Guo, Feng-Kun and Hanhart, Christoph and Mei\ss{}ner, Ulf-G. and Wang, Qian and Zhao, Qiang and Zou, Bing-Song",
    title = "{Hadronic molecules}",
    eprint = "1705.00141",
    archivePrefix = "arXiv",
    primaryClass = "hep-ph",
    doi = "10.1103/RevModPhys.90.015004",
    journal = "Rev. Mod. Phys.",
    volume = "90",
    number = "1",
    pages = "015004",
    year = "2018",
    note = "[Erratum: Rev.Mod.Phys. 94, 029901 (2022)]"
}

@article{Meissner:2020khl,
    author = "Mei\ss{}ner, Ulf-G.",
    title = "{Two-pole structures in QCD: Facts, not fantasy!}",
    eprint = "2005.06909",
    archivePrefix = "arXiv",
    primaryClass = "hep-ph",
    doi = "10.3390/sym12060981",
    journal = "Symmetry",
    volume = "12",
    number = "6",
    pages = "981",
    year = "2020"
}
%%%%%%%%%%%%%%%%%%%%%%%%%%%%
%%%%%%%%%%%%%%%%%%%%%%%%%%%%

\clearpage
%%%%%%%%%%%%%%%%%%%%%%%%%%%%%%%%%%
%%%%%%%%%%%%%%%%%%%%%%%%%%%%%%%%

\end{document}